\documentclass[aps,pra,twocolumn,amsmath,amssymb,10pt,showpacs]{revtex4-1}
\usepackage{graphicx}

\usepackage{color}

\newcommand{\EN}{\ensuremath{E_\mathcal{N}}}

\newcommand{\bk}{\ensuremath{\mathbf k}}
\newcommand{\br}{\ensuremath{\mathbf r}}

\begin{document}

\title{Entanglement of remote quantum systems by environmental modes}
\author{Friedemann Queisser$^1$, Thomas Zell$^2$, and Rochus Klesse$^2$}
\affiliation{$^1$Fakult\"at f\"ur Physik, Universit\"at Duisburg-Essen, Lotharstrasse 1, D-47057 Duisburg, Germany\\
$^2$Universit\"at zu K\"oln, Institut f\"ur Theoretische Physik 
  Z\"ulpicher Str. 77, D-50937 K\"oln, Germany}
\date{\today}

\begin{abstract}
We investigate the generation of quantum mechanical entanglement 
of two remote oscillators that are locally coupled to a common bosonic 
bath. Starting with a Lagrangian formulation of a suitable model, we 
derive two coupled Quantum Langevin Equations that exactly describe the 
time evolution of the two local oscillators in presence of the coupling to 
the bosonic bath. Numerically obtained solutions of the Langevin Equations 
allow us to study the entanglement generation between the oscillators in 
terms of the time evolution of the logarithmic negativity. Our results 
confirm and extend our previously obtained findings, namely that 
significant entanglement between oscillators embedded in a free bosonic 
bath can only be achieved if the system are within a microscopic distance. 
We also consider the case where the bosonic spectral density is 
substantially modified by imposing boundary conditions on the bath modes. 
For boundary conditions corresponding to a wave-guide like geometry of the 
bath we find significantly enlarged entanglement generation. This 
phenomenon is additionally illustrated within an approximative model that 
allows for an analytical treatment.
\end{abstract}


\maketitle

\section{Introduction}\label{sec-introduction}\label{sec0}   
Consider two remote microscopic quantum-mechanical objects that are
locally coupled to a common bosonic heat bath. It is assumed
that there are no direct interactions between the microscopic objects. In
this situation any entanglement of the two objects tends to rather rapidly
decline with time, as it is described by the general phenomenon of 
quantum-mechanical decoherence \cite{JZKGKS03,S07}. On the other hand, by
exchanging bath 
bosons the objects do indirectly interact with each other, which, as
like any interaction, tends to entangle the two objects. 
Depending on which of the two effects of the bath is dominant, the two
objects in an initially separable common state will either stay
separable or develop some amount of entanglement. 
To find out whether under some given conditions the former or the
latter case is realized is generally a difficult theoretical problem
which has been addressed in numerous publications during the last
decade \cite{Bra02,BFP03,Bra05,BF06,OK06,Pra04,AZ07,CYH08,LG07,HB08,PR08,STP07}. 

While there seems to be agreement that for sufficiently small distances
entanglement develops under quite general conditions, it is still
controversial whether the mechanism under discussion 
can also generate entanglement between objects in a macroscopic distance. 
Here we extend and generalize our previous work on the distance
dependence of entanglement generation via a bosonic heat bath \cite{ZQK09}.  
Our main finding is that for a heat bath exhibiting a continuum of low
frequency modes dissipative couplings will always limit the range
within which entanglement can be generated to very short distances.

In Sec.~\ref{sec1} we formulate a model system consisting of two harmonic
oscillators that are locally and linearly coupled to a bosonic heat
bath. We use a positive definite Lagrangian which generalizes the one
that has been used by Unruh and Zurek in a similar context
\cite{UZ89}. After transforming to a Hamiltonian description we end
up with 
a Hamilton operator that includes precisely the non-local counter
term that we motivated by a different reasoning in \cite{ZQK09}.
While in \cite{ZQK09} we used spectral coupling density of states of
Drude type, here we consider spectral densities
$J(\omega)$ with exponential cut-off at high frequencies 
and power-law behaviour $J(\omega)\sim \omega^s$ with $s=1$ and $3$ at
low frequencies. The bath dimension is assumed to be either one or three.

We derive a set of two coupled quantum-Langevin
equation that describes the time evolution of the two local
oscillators. From these equations we obtain the time evolution
of the entanglement of the two oscillators, measured in logarithmic
negativity, essentially by use of numerical Laplace
transformation. Here we employ an efficient new numerical method
\cite{WW08} that allows for a better exploration of parameter space
(Sec.~\ref{sec2}).

Strongly non-resonant frequency modes are the main source of
decoherence, while modes with frequency near oscillator (spin)
frequency $\Omega_0$ tend to entangle the two distant oscillators.
Hence, suppressing the former and at the same time enhancing the latter
should increase the entanglement of the oscillators. 
An extreme case of this type is obviously realized by a single-mode cavity
that is coupled to two systems with resonant energy levels. It
is known that in this system entanglement can be efficiently 
generated. 
A variant of such a system that might be useful for entanglement over wider
distances is a tube, i.e. a long cavity. 
By the boundary conditions imposed on transversal modes, here the spectrum 
exhibits a gap ranging from zero frequency to some finite $\omega_g > 0$.
We analyze the entanglement
generation of these type of systems first with the
numerical methods that we already used before in Sec.~\ref{sec4}, 
and then analytically within a simplified model system in
Sec.~\ref{sec5}.
In comparison to a free heat bath, we find for the tube 
significant enlarged entanglement. 
We conclude with a summary and a discussion of our results in
Sec.~\ref{sec6}.

\section{Model}\label{sec1}

\subsection{Lagrangian}\label{subsec-model}
The model within we investigate bath mediated generation of
entanglement consists of two identical oscillators of frequency
$\Omega_0$ located at positions $\pm\mathbf{r}/2$ 
and locally coupled to an otherwise free scalar field
$\phi(\mathbf x)$.
The two oscillators represent the two remote microscopic system and
are referred to as  {\em system oscillators} in the following. 
Their canonical variables are denoted by $Q_{1/2}$ and $P_{1/2}$,
their masses are set to
unity. In order to avoid runaway solutions we choose a bilinear
coupling to the velocity of the  
scalar field \cite{UZ89} and thus arrive at a Lagrangian

 \begin{eqnarray}
 L&=&\frac{1}{2}\sum_{i=1,2}\left(\dot{Q}_i^2-\Omega_0^2Q_i^2\right)+\frac{1}{2}\int
 d^Dx\left(\dot{\phi}^2-(\nabla \phi)^2\right)\nonumber\\ 
 &+&\int d^Dx\:\dot{\phi}(\mathbf{x}) \left( g\left(\mathbf{x}-\frac{\mathbf{r}}{2}\right) Q_1
   +g\left(\mathbf{x}+\frac{\mathbf{r}}{2}\right)Q_2\right)\,. 
 \label{eq-lagragian}
 \end{eqnarray}
The coupling function $g(\mathbf{x})$ is assumed to be peaked  around
0 and to have a width of order $\delta x$, describing the 
minimal distance which can be resolved by the bath modes $\phi(\mathbf
x)$. 

Quantizing the model leads to the positive definite Hamiltonian
\begin{eqnarray*}
\hat{H}&=&\frac{1}{2}\int d^D x\bigg[\hat{\Pi}_\phi-g \left(\mathbf{x}-
\frac{\mathbf{r}}{2}\right)\,\hat{Q}_1 -g
\left(\mathbf{x}+\frac{\mathbf{r}}{2}\right)\,\hat{Q}_2
\bigg]^2\nonumber\\ 
& &+\frac{1}{2}\int d^D x(\nabla\hat{\phi})^2+
\frac{1}{2}\sum_{i=1,2}\left(\hat{P}_i^2+\Omega_0^2\hat{Q}_i^2\right)\,. 
\end{eqnarray*}
Confining the field within a large $D$-dimensional box of size $l \gg r$ and
imposing periodic boundary conditions the field operators 
$\hat\phi(\mathbf x)$, $\hat\Pi(\mathbf x)$, and the coupling function
$g(\mathbf x)$ can be decomposed in Fourier components as
\begin{eqnarray*}
\hat{\phi}(\mathbf{x})&=&\frac{1}{l^{D/2}}\sum_\mathbf{k}\hat{\phi}_\mathbf{k}e^{i\mathbf{kx}}\,,\\ 
\hat{\Pi}_\phi(\mathbf{x})&=&\frac{1}{l^{D/2}}\sum_\mathbf{k}\hat{\Pi}_\mathbf{k}e^{i\mathbf{kx}}\,,\\
g(\mathbf{x})&=&\frac{1}{l^{D/2}}\sum_\mathbf{k}g_{\mathbf{k}}e^{i\mathbf{kx}}\: .
\end{eqnarray*}
Expressed with these modes, the Hamiltonian reads
\begin{eqnarray}\label{Hamilt}
\hat{H}&=&\frac{1}{2}\sum_\mathbf{k}\left(\hat{\Pi}_\mathbf{k}\hat{\Pi}_\mathbf{-k}+k^2\hat{\phi}_\mathbf{k}\hat{\phi}_\mathbf{-k}\right)+
\frac{1}{2}\sum_i\left(\hat{P}_i^2+\Omega_0^2 \hat{Q}_i^2\right)\nonumber\\
& &-\sum_\mathbf{k}g_{-\mathbf{k}}\hat{\Pi}_\mathbf{k}\left(e^{\frac{i}{2}\mathbf{kr}}\hat{Q}_1+e^{-\frac{i}{2}\mathbf{kr}}\hat{Q}_2\right)\nonumber\\
& &+\frac{1}{2}\sum_\mathbf{k}g_{-\mathbf{k}}g_{\mathbf{k}}\left(\hat{Q}_1^2+2\hat{Q}_1\hat{Q}_2e^{i\mathbf{k r}}+\hat{Q}_2^2\right)\,.
\end{eqnarray}
The appearance of the operator product $\hat Q_1 \hat Q_2$
in the last term of the Hamiltonian does {\em not} indicate a direct
coupling of the system oscillators. It is merely a simple consequence
of transforming from the Lagrangian formalism (in which the model
evidently {\em is} local, cf.~(\ref{eq-lagragian})) to the Hamiltonian 
formalism. In our previous work \cite{ZQK09} we directly formulated a
model in the Hamiltonian formalism, which made it necessary
to add by hand an appropriate counter-term in order to ensure
stability and causality of the resulting dynamics. In fact, the
resulting Hamiltonian in \cite{ZQK09} agrees with the present one up
to an irrelevant change of oscillator variables $Q_i\leftrightarrow P_i$.

\subsection{Equations of motion}\label{subsec-equations_of_motion}
It is a straightforward task to derive the Heisenberg equations of motion
for the environmental modes and the system oscillators,
\begin{eqnarray}
\dot{\hat{ Q}}_{1/2}&=&\hat {P}_{1/2}\label{oszi1}\,,\\
\dot{\hat{ P}}_{1/2}&=&-\Omega_0^2\hat{Q}_{1/2}+\sum_\mathbf{k}g_{-\mathbf{k}}\hat{\Pi}_\mathbf{k}e^{\pm \frac{i}{2} \mathbf{kr}}\nonumber\\
& &-
\sum_\mathbf{k}g_\mathbf{k}g_{-\mathbf{k}}(\hat{Q}_{1/2}+\hat{Q}_{2/1}e^{i\mathbf{kr}})\label{oszi2}\,,\\
\dot{\hat{\phi}}_\mathbf{k}&=&\hat{\Pi}_\mathbf{k}-g_{\mathbf{k}}\left(e^{-\frac{i}{2}\mathbf{kr}}\hat{Q}_1
+e^{\frac{i}{2}\mathbf{kr}}\hat{Q}_2\right)\label{diffphi1}\,,\\
\dot{\hat{\Pi}}_\mathbf{k}&=&-k^2\hat{\phi}_\mathbf{k}\label{diffphi2}\:.
\end{eqnarray}

In our analysis presented below we always assume that at time $t=0$
the system oscillators are prepared in some initial state, independent
from the environmental state. For $t\ge 0$ both system oscillators and 
environment evolve according to the Lagrangian
Eq.~(\ref{eq-lagragian}) or equivalently according to
Eqs.~(\ref{oszi1}) to (\ref{diffphi2}). For $t<0$ the environment is
assumed to evolve freely, without coupling to the system oscillators. 
We therefore require $\hat \phi_\mathbf{k}(t)$ and $\hat
\pi_\mathbf{k}(t)$ to be solutions of Eqs.~(\ref{oszi1}) to (\ref{diffphi2}) 
for $t\ge 0$ and to agree with
\begin{eqnarray*}
\hat{\phi}_{\mathbf{k},\mathrm{hom}}(t)&=&\frac{1}{\sqrt{2k}}
\left(\hat{a}_\mathbf{k}e^{-ikt}+\hat{a}^\dagger_\mathbf{-k}e^{ikt}\right)\,\\
\hat{\Pi}_{\mathbf{k},\mathrm{hom}}(t)&=&-i\sqrt{\frac{k}{2}}
\left(\hat{a}_\mathbf{k}e^{-ikt}-\hat{a}^\dagger_\mathbf{-k}e^{ikt}\right)\,,
\end{eqnarray*}
for $t<0$. Here $\hat{a}_\mathbf{k}^\dagger$ and $\hat{a}_\mathbf{k}$
denote creation and annihilation operators of the field mode of
wavevector $\mathbf k$. 

It can be easily checked that these field operators are given by
\begin{eqnarray}
\hat{\phi}_\mathbf{k}(t)&=&\hat{\phi}_{\mathbf{k},\mathrm{hom}}(t)\nonumber\\
& &\hspace{-1cm}-g_{k}\int_0^t dt' \cos(k(t-t'))
\left(e^{-\frac{i}{2}\mathbf{kr}}\hat{Q}_1(t')+e^{\frac{i}{2}\mathbf{kr}}\hat{Q}_2(t')\right)\nonumber\\
\hat{\Pi}_\mathbf{k}(t)&=&\hat{\Pi}_{\mathbf{k},\mathrm{hom}}(t)\label{fieldequ2}\\
& &\hspace{-1cm}+g_\mathbf{k}\int_0^t dt' k \sin(k(t-t'))  
\left(e^{-\frac{i}{2}\mathbf{kr}} \hat{Q}_1(t')+e^{\frac{i}{2}
    \mathbf{kr}}\hat{Q}_2(t')\right) \nonumber\,.         
\end{eqnarray}

\subsection{Quantum Langevin Equations}\label{subsec-quantum_langevin_equations}
Inserting these expressions into equations (\ref{oszi1}) and
(\ref{oszi2}), one obtains two Quantum Langevin Equations (QLEs)
for the dynamics of the system operators $\hat Q_i(t)$ for $t\ge 0$,
\begin{eqnarray} \label{eq-dynamics}
\ddot{\hat Q}_i(t) + \Omega_0^2 \hat Q_i(t) + \frac{d}{dt} \int_0^t
dt' &[ &
\Gamma_0(t-t')\hat Q_i(t') \\
&+& \Gamma_r(t-t') \hat Q_{\bar i}(t')] = \hat B_i(t)\:, \nonumber
\end{eqnarray}
where $(i,\bar i)= (1,2)$ and $(2,1)$. Here, we  
introduced a damping kernel
\begin{equation*}
\Gamma(\mathbf r, t)=\sum_{\mathbf{k}}g_\mathbf{k}g_\mathbf{-k}\cos(kt) e^{i\mathbf{kr}}\:,
\end{equation*}
and generalized forces given by bath operators 
\begin{equation*}
\hat B_{1/2}(t) = \sum_\mathbf{k}g_{-\mathbf{k}}\hat\Pi_{\mathbf{k},\mathrm{hom}}(t)
e^{\pm i\frac{\mathbf{kr}}{2}}\:.
\end{equation*}
It is worth emphasizing that these bath operators evolve freely in
time; the back-action of the two system oscillators is solely contained
in the memory terms of Eqs. (\ref{eq-dynamics}).

The two QLEs can be written in a more convenient form if we combine the variables of
the system oscillators in a vector  
\begin{eqnarray}
\hat{\mathbf{y}}(t)&=&(\hat{Q}_1,\hat{Q}_2,\hat{P}_1,\hat{P}_2)^T \:,
\end{eqnarray}
define a generalized force vector by
\begin{eqnarray}
\hat{\mathbf{B}}(t)&=& (0,0,B_1(t), B_2(t)\:)\:,
\end{eqnarray}
a mass-frequency matrix
\begin{eqnarray}
\mathcal{Z}&=&
\begin{pmatrix}
0&0&-1&0\\
0&0&0&-1\\
\Omega_0^2&0&0&0\\
0&\Omega_0^2&0&0
\end{pmatrix}\,,\\
\end{eqnarray}
and, finally, a matrix
\begin{eqnarray}
\mathcal{C}(t)&=&\begin{pmatrix}
0&0&0&0\\
0&0&0&0\\
\Gamma(\mathbf 0, t)&\Gamma(\mathbf r, t)&0&0\\
\Gamma(\mathbf r, t)&\Gamma(\mathbf 0,t)&0&0
\end{pmatrix}\: .
\end{eqnarray}
With these definitions the QLE states
\begin{eqnarray}\label{eq:QLE}
\dot{\hat{\mathbf{y}}}(t)+\mathcal{Z}\hat{\mathbf{y}}(t)+\frac{d}{dt}\int_0^t
dt'\mathcal{C}(t-t' )
\hat{\mathbf{y}}(t')=\hat{\mathbf{B}}(t)\,.
\end{eqnarray}
Formally, its solution $\hat{\mathbf y}(t)$ for initial $\hat{\mathbf
  y}(0)$ and force $\hat{\mathbf B}(t)$ is 
\begin{eqnarray}\label{eq-solution}
\hat{\mathbf{y}}(t)=\mathcal{G}(t)\hat{\mathbf{y}}(0)-\int_0^t dt' \mathcal{G}(t-t')\hat{\mathbf{B}}(t')\,,
\end{eqnarray}
where $\mathcal{G}(t)$ is Green's function of the QLE (\ref{eq:QLE}). 

\subsection{Covariance matrix and logarithmic negativity}\label{subsec-covariance}
Correlations and entanglement of the system oscillators can be studied
on the basis of the covariance matrix
\begin{eqnarray*}
\mathrm{Cov}_{lm}(t)=\mathrm{Tr}_\mathrm{S}[\{\hat{y}_l(t),\hat{y}_m(t)\}\hat{\rho}_\mathrm{S}(0)]\,, 
\end{eqnarray*} 
where $\hat{\rho}_\mathrm{S}(t)$ denotes the density matrix of the system oscillators.
Demanding that system and environment are initially in a factorizing state, 
$\hat{\rho}_\mathrm{SB}(0)=\hat{\rho}_\mathrm{S}(0)\otimes\hat{\rho}_\mathrm{B}$,
the time evolution of the covariance matrix can be expressed via
Eq. (\ref{eq-solution}) and a bath correlation matrix 
\begin{equation*}
\mathcal{K}(t)=\mathrm{Tr}_\mathrm{B}[\{\mathbf{B}(t),\mathbf{B}(0)^\dagger\}\hat{\rho}_\mathrm{B}]
\equiv 
\begin{pmatrix}
0&0&0&0\\
0&0&0&0\\
K(\mathbf 0, t)&K(\mathbf r, t)&0&0\\
K(\mathbf r, t)&K(\mathbf 0,t)&0&0
\end{pmatrix}
\end{equation*}
as 
\begin{eqnarray}\label{eq-covmatrix}
\mathrm{Cov}(t)&=&\mathcal{G}(t)\mathrm{Cov}(0)\mathcal{G}(t)^\dagger\\
& &+\int_0^t dt'\int_0^t dt''\mathcal{G}(t'-t)\mathcal{K}(t'-t'')\mathcal{G}(t-t'')^\dagger\nonumber\,.
\end{eqnarray}
Throughout the paper we will restrict ourselves to the consideration
of Gaussian states which are entirely determined by  
their covariance matrix. 
Then, a convenient measure for the entanglement
is the logarithmic negativity 
\begin{eqnarray*}
E_\mathcal{N}=-\sum_{i=1,2}\log_2\left(\mathrm{min}(1,\lambda_i)\right)\,,
\end{eqnarray*}
where $\lambda_{1/2}$ are the symplectic eigenvalues of the partial time-reversed
covariance matrix $\mathrm{Cov}^{T_B}=P\mathrm{Cov}P$ with $P=\mathrm{diag}(1,1,1,-1)$. 
Loosely speaking, the logarithmic negativity measures the deviation
of the oscillators being in a separable state \cite{VW02}.

\subsection{Spectral coupling density of  states, thermal bath correlations}\label{subsec-spectral}
The model needs to be further specified by fixing the oscillator-bath
couplings $g_\mathbf k$. To this end we assume isotropy,
i.e. $g_\mathbf k = g_k$, and define a spectral coupling density as
\begin{equation}
J(\omega) =\sum_\mathbf k |g_\mathbf k|^2 \omega \delta(\omega - k)\:,
\end{equation}
which in turn we suppose to be of the form 
\begin{equation}\label{eq:specdens}
J(\omega) = \frac{8 \gamma}{\pi} \omega \left(
  \frac{\omega}{\Omega_c} \right)^{s-1} e^{- \omega/\Omega_c}\:.
\end{equation}
Here, $\gamma$ denotes the overall coupling strength, $s$ is the
spectral index, and $\Omega_c$ is a cutoff frequency. 
Note that the definition of the spectral coupling density differs from
the usual definition in the extra $\omega$-factor, which attributes to
the fact that we consider a coupling to the field velocities rather
than to the field itself.

With this definitions we obtain for the damping kernels the explicit 
expressions
\begin{eqnarray}\label{eq:damping-kernel}
 \Gamma(\mathbf r,t)=\begin{cases}
            \int_0^\infty d\omega\frac{J^{1D}(\omega)}{\omega}\cos(\omega t)\cos(\omega r)  \\
 	    \int_0^\infty d\omega\frac{J^{3D}(\omega)}{\omega}\cos(\omega t)\frac{\sin(\omega r)}{\omega r} 
             \end{cases}
 \end{eqnarray}
for a one-dimension and a three-dimensional bath, respectively.
When the environment initially is in a thermal state $\hat \rho_B$ of
temperature $T$, the bath correlations determines to 
\begin{eqnarray}\label{eq:bath-correlation}
K(\mathbf r,t)=\begin{cases}
            \int_0^\infty d\omega
           J^{1D}(\omega)\coth\left(\frac{\omega}{2T}\right)\cos(\omega
           t)\cos(\omega r)  \\ 
	     \int_0^\infty d\omega J^{3D}(\omega)\coth\left(\frac{\omega}{2T}\right)\cos(\omega t)\frac{\sin(\omega r)}{\omega r}\,.
            \end{cases}
\end{eqnarray}

\section{Numerical Solutions}

\subsection{Methods}

The solution of the integro-differential equations (\ref{eq-dynamics}) which is 
formally given by the expression (\ref{eq-solution}) was obtained numerically.

In order to solve the homogeneous part (\ref{eq-dynamics}) in the position
basis we introduce the auxiliary functions \cite{WW08}
\begin{eqnarray*}
f_{i}({\mathbf r},t,u)&=&\int_0^t dt' \big(\Gamma({\mathbf 0},t-t'+u)Q_i(t')\\
& &\hspace{2.5cm}+\Gamma({\mathbf r},t-t'+u)Q_{\bar{i}}(t')\big)
\end{eqnarray*}
which satisfy the partial differential equations
\begin{eqnarray*}
\partial_t f_i({\mathbf r},t,u)=\Gamma({\mathbf 0},u)Q_i(t)+\Gamma({\mathbf r},u)Q_{\bar{i}}(t)+\partial_u f_{i}({\mathbf r},t,u)\,.
\end{eqnarray*}
By means of the Fourier transforms 
\begin{eqnarray*}
f_i({\mathbf r},t,u)&=&\frac{1}{\sqrt{2\pi}}\int_{-\infty}^\infty ds\, e^{ius}\hat{f}_i({\mathbf r},t,s)\\
\Gamma({\mathbf r},u)&=&\frac{1}{\sqrt{2\pi}}\int_{-\infty}^\infty ds\, e^{ius}\hat{\Gamma}({\mathbf r},s)\\
\end{eqnarray*}
we obtain the ordinary differential equations
\begin{eqnarray}\label{eq-diffsys1}
\partial_t \hat{f}_i({\mathbf r},t,s)&=&\hat{\Gamma}({\mathbf 0},s)Q_i(t)+\hat{\Gamma}({\mathbf r},s)Q_{\bar{i}}(t)\nonumber\\
& &\hspace{2.5cm}+is \hat{f}_i({\mathbf r},t,s)\,.
\end{eqnarray}
Discretizing the inverse Fourier transform of $\tilde{f}(t,s)$ we can rewrite the homogeneous part of (\ref{eq-dynamics})
according to
\begin{eqnarray}\label{eq-diffsys2}
\ddot{Q}_i(t)+\Omega_0^2 Q_i(t)=-\sum_{k=1}^{2n_\mathrm{grid}+1} \frac{\triangle s_k}{\sqrt{2\pi}}\partial_t \hat{f}_i({\mathbf r},t,s_k)\,.
\end{eqnarray}
where $\triangle s_k$ denotes the grid spacing.
Equations (\ref{eq-diffsys1}) and (\ref{eq-diffsys2}) form a coupled system of ordinary 
differential equations which can be solved by standard algorithms.

Using the numerical solutions for $Q_{i}$ and $P_{i}$, the Greens function can be deduced from 
the homogeneous part of (\ref{eq-solution}).
Futhermore, in order to obtain the covariance matrix (\ref{eq-covmatrix}), three numerical 
integrations have to be nested, two for the time integrations and one for the
bath correlator (\ref{eq:bath-correlation}) which is not available in closed analytical form.

\subsection{Free Space Environment}
\label{sec2}

We now present our results from the exact numerical solution of our
model.
For the free bath, which is discussed in the following section, we used a 
grid size of $n_\mathrm{grid}=10000$ and constant grid spacing
$\triangle s_k$ with $s_\mathrm{max}=(2n_\mathrm{grid}+1)\triangle s_k=10\Omega_c$.
As we want to study the creation of entanglement between the
two system oscillators, we start with a separable state
$\rho_S(0)$. For simplicity, $\rho_S(0)$ will always be the ground
state in the following.
The bath is initialized in a thermal state $\rho_T$ with zero temperature.

We look at both the asymptotic values reached after the system has
equilibrated, as well as the dynamics for short and intermediate
times. The free parameters, which remain after scaling out mass and
frequency of the oscillators, are distance $r$, cut-off frequency
$\Omega_c$, and damping $\gamma$. Generally, we measure distances in
units of $c/\Omega_0$ and frequencies in units of $\Omega_0$.

\begin{figure}
  \centering
  \includegraphics{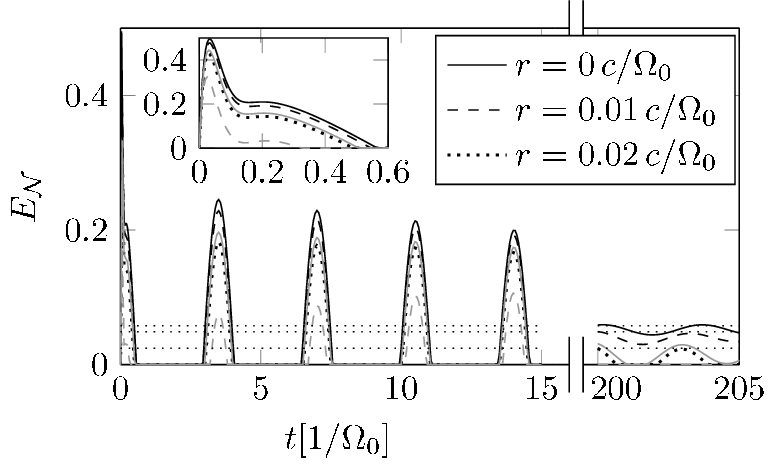}
  \caption{Logarithmic negativity $E$ as function of time (in units of
    $1/\Omega_0$) for different distances for the 1D bath (in gray)
    and the 3D bath (in black). Dashed lines represent asymptotic
    values. The parameters are the same as in
    Fig.~\ref{fig:ln_asym}. A zoomed in view of the beginning of the
    evolution $t < 0.6/\Omega_0$ is shown in the inset.}
  \label{fig:ln_dyn}
\end{figure}

The full time evolution exhibits fast dynamics at the beginning due to
the coupling of the bath being switched on suddenly, see Fig.~\ref{fig:ln_dyn}. 
Later the entanglement oscillates until it reaches its asymptotic
value.
For the physically unrealistic case $r = 0$, the 
relative position $Q_2 - Q_1$ remains undamped during the whole time evolution. 
The damping increases for larger distances $r$. 
From Fig.~\ref{fig:ln_dyn} it can be deduced that the asymptotic value is
reached more quickly for larger distances.

\begin{figure}
  \centering
  \includegraphics{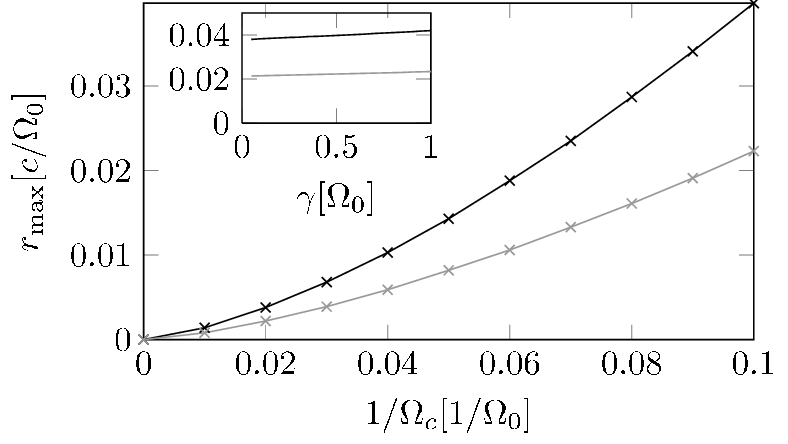}
  \caption{Smallest distance $r_\text{max}$ at which the oscillators
    stay separable during the entire time evolution as a function of the inverse cut-off $1/\Omega_c$ for
    the 1D bath (in gray) and the 3D bath (in black). 
    The parameters
    are the same as in Fig.~\ref{fig:ln_asym}. The inset shows
    $r_\text{max}$ as function of the coupling strength $\gamma$.}
  \label{fig:ln_d_O}
\end{figure}

As the maximum entanglement reached at short times is larger than the
asymptotic value, one can ask the question if it features a more
favorable distance dependence. We have plotted the smallest distance
at which the oscillators stay separable during the entire time
evolution as a function of the inverse cut-off in
Fig.~\ref{fig:ln_d_O}. 
While the distance is somewhat larger than in
the asymptotic case, see Fig.~\ref{fig:ln_d_asym}, it can still be
upper bounded by a linear function proportional to
$1/\Omega_c$. Therefore the qualitative behavior is the same. Exactly as
in the asymptotic case, the coupling strength $\gamma$ only has a
minor influence on $r_\text{max}$.

\begin{figure}
  \centering
  \includegraphics{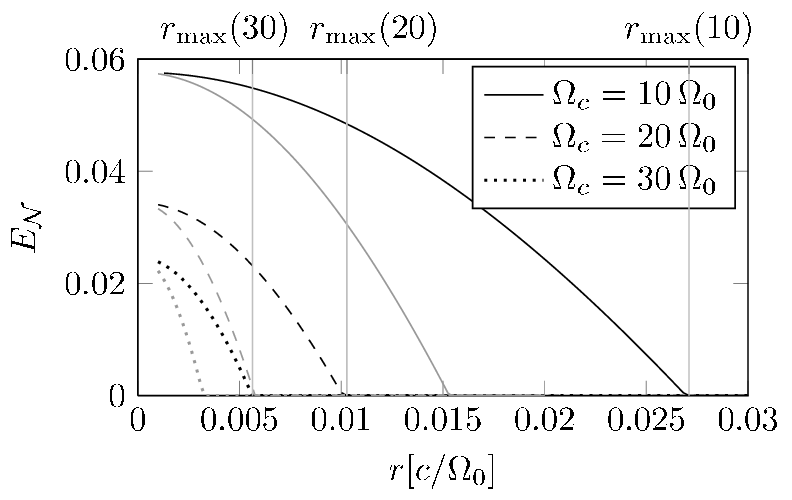}
  \caption{Asymptotic entanglement of the system oscillators measured
    in logarithmic negativity $E$ as a function of distance $r$ (in
    units of $c/\omega_0$) for the 1D bath (in gray) and the 3D bath
    (in black), damping constant $\gamma =\Omega_0$, and cut-off
    frequency $\Omega_c = 10 \Omega_0$.  $E$ drops to zero at a rather
    small critical distance $r_\text{max}\lesssim c/\Omega_c$
    (cf. Fig.~\ref{fig:ln_d_asym}).}
  \label{fig:ln_asym}
\end{figure}

\begin{figure}
  \centering
  \includegraphics{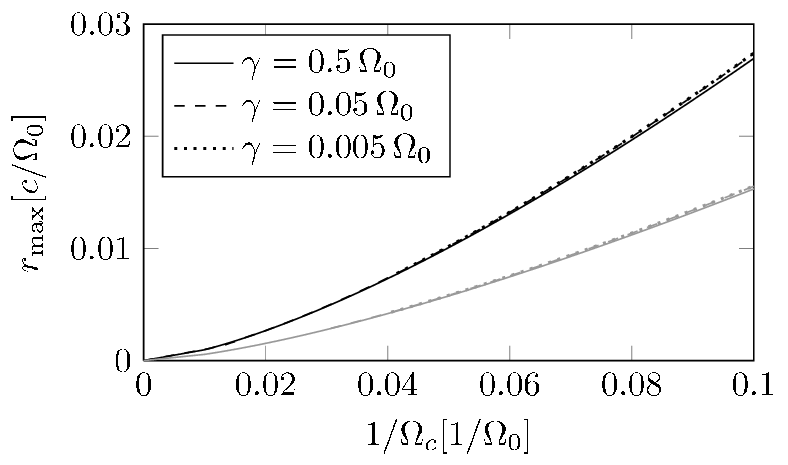}
  \caption{Distance $r_\text{max}$ at which the oscillators become
    asymptotically separable for the 1D bath (in gray) and the 3D bath
    (in black). The parameters are the same as in
    Fig.~\ref{fig:ln_asym}.}
  \label{fig:ln_d_asym}
\end{figure}

The asymptotic state can be calculated by taking the Fourier transform
of the Langevin equation~(\ref{eq:QLE})~\cite{GZ04}, from which we
have dropped the quickly decaying term $\mathbf y(0) \mathcal
C(t)$. The resulting Quantum Langevin equation
\begin{equation*}
  \dot{\mathbf y}(t) + \mathcal Z \mathbf y(t) + \int_{-\infty}^t dt'\,
  \theta(t-t')\mathcal C(t-t')\mathbf{\dot y}(t') = \mathbf B(t)
\end{equation*}
is transformed into the algebraic equation
\begin{equation*}
  -i\omega\hat{\mathbf y}(\omega) + \mathcal Z \hat{\mathbf y}(\omega) -i\omega \mathcal
  R(\omega) \hat{\mathbf y}(\omega) = \mathbf B(\omega)
\end{equation*}
with the matrix $\mathcal R(\omega)$ defined by
\begin{align*}
  \tilde{\mathcal C}(\omega) &= \frac{1}{2} \int_{-\infty}^\infty dt \,
  e^{i\omega t} \mathcal C(t)\\
  \mathcal R(\omega) &= \int_0^\infty dt\, e^{i \omega t} \mathcal C(t)
  = \tilde{\mathcal C}(\omega) +  \frac{i}{\pi} P \int_{-\infty}^\infty d\omega'
  \frac{\tilde{\mathcal C}(\omega')}{\omega'-\omega}\,.
\end{align*}
The Fourier transformation of $\mathcal C(t)$ cancels with the
integration over $\omega$, so that the matrix elements
$\tilde{\Gamma}(\mathbf{r},\omega)$ of $\tilde{\mathcal C}(\omega)$ corresponding to the
elements $\Gamma(\mathbf{r},t)$ of $\mathcal C(t)$ are simply given by
\begin{equation*}
  \tilde{\Gamma}(\mathbf{r},\omega) = 
  \frac{\pi J(\left|\omega\right|)}{4\left|\omega\right|} \cos(\omega r)\,.
\end{equation*}
The equilibrium equal time correlation function of the system can then
be expressed in terms of the matrix
\begin{equation*}
  \mathcal F(\omega) = (-i \omega + \mathcal Z -i\omega \mathcal R(\omega))^{-1}
\end{equation*}
as the integral
\begin{multline*}
  \langle\lbrace \mathbf{ y}_i(0), \mathbf{y}_j(0)
  \rbrace\rangle =
  \frac{1}{\Omega_0} \cdot\\
  \sum_{k,l \in \lbrace 3,4 \rbrace} \int_{-\infty}^\infty d\omega \,
  \mathcal F_{ik}(\omega) \mathcal F_{jl}(-\omega)
  J(\left|\omega\right|) \cos(\omega r \delta_{kl})\,.
\end{multline*}
The behavior of the asymptotic entanglement as a function of distance
$r$ is plotted in Fig.~\ref{fig:ln_asym}. It drops to zero for
rather small distances $r_\text{max}$, which are on the order of the
inverse cut-off $1/\Omega_c$ as shown in
Fig.~\ref{fig:ln_d_asym}. This is only very weakly dependent on the
coupling strength, which can be explained by the fact that $\gamma$
controls both coupling and decoherence at the same time. Changing it
cannot be used to increase the entanglement capabilities of the bath.

\section{Waveguide environment}
\label{sec4}
While environmentally created entanglement does not seem to be
possible in free space, changing the geometry can have a big impact. A
simple idea is to make the bath effectively one-dimensional by placing
the two oscillators inside a waveguide with quadratic cross-section of
side length $a$. The quantization in transverse direction suppresses
all frequencies below the first mode with frequency $\omega_0 =
\sqrt{8}\pi /a$. In order to keep things simple, we will for now only
consider this first mode in transverse direction. The spectrum in
longitudinal direction will remain continuous.

We split the wave vector $\bk$ into its components
\begin{equation*}
  \bk = \frac{2\pi n_1}{l} \mathbf e_1 + \frac{2\pi}{a}(n_2
  \mathbf e_2 + n_3 \mathbf e_3)\,.
\end{equation*}
with  $n_i \in \mathbb{Z} \setminus \lbrace 0\rbrace$.

The component in $\mathbf e_1$-direction can approximated by an integral due to the
large length of the tube in this direction. For this, we make a change
of variables $n_1 \rightarrow \left|\bk\right| = k$ which
implies
\begin{align*}
  dn_1 = \frac{l}{2\pi} \frac{k\,dk}{\sqrt{k^2 -
      \frac{\omega_0^2}{2}(n_2^2 + n_3^2)}}\,.
\end{align*}
At the same time we will examine what happens to an additional term
$e^{i\bk\br}$ as it occurs in the damping
kernel~\eqref{eq:damping-kernel} and in the bath
correlator~\eqref{eq:bath-correlation}. 
Assuming that $\br \parallel \mathbf e_1$ we find for the damping kernel
\begin{eqnarray}\label{eq:damping-kernel2}
\Gamma({\mathbf r},t)&=&\int_0^\infty d\omega \sum_\bk\left|g_\bk\right|^2
  \delta(\omega - k)\, e^{i \bk \br}\\
  & &\hspace{-2cm}=\int_0^\infty d\omega  \sum_{n_2, n_3 = 1}^{\frac{\omega_0^2 }{2}(n_2^2 + n_3^2) <
    \omega^2}\frac{8l}{2\pi}\omega \left|g_\omega\right|^2\cos(\omega t)\nonumber\\
& &\hspace{1cm}\times\frac{\cos\left(r \sqrt{\omega^2 -
  \frac{\omega_0^2}{2}(n_2^2 + n_3^2)}\right)}{\sqrt{\omega^2 -
  \frac{\omega_0^2}{2}(n_2^2 + n_3^2)}}\nonumber
\end{eqnarray}
where the factor $8$ is the number of octants of a sphere. 
Taking in (\ref{eq:damping-kernel2}) the limit $\omega_0\rightarrow0$ and comparing
with (\ref{eq:damping-kernel}) we find 
\begin{equation*}
  \left|g_\omega\right|^2 = \frac{J(\omega)\omega_0^2}{4l\omega^3}\,,
\end{equation*}
where we assumed isotropy of the spectral couplings, that is $g_{\mathbf k}=g_k$.
If the oscillator frequency $\Omega_0$ is close to the first
excitation, we can approximate the sum in (\ref{eq:damping-kernel2}) by the
first term $n_2=n_3 = 1$.
Thus, the damping kernel in the waveguide reads
\begin{equation*}
  \Gamma_\mathrm{wg}({\mathbf r},t) =  \int_0^\infty d\omega\, \frac{J_\mathrm{wg}(\omega)}{\omega}
  \cos(\omega t) \cos\left(r \sqrt{\omega^2 - \omega_0^2}\right)
\end{equation*}
with
\begin{equation}\label{eq:vanHovespec}
  J_\mathrm{wg}(\omega) = \frac{8 \gamma
    \omega_0^2}{\pi^2 \sqrt{\omega^2 - \omega_0^2 }}
  \left(\frac{\omega}{\Omega_c}\right)^{s-1} e^{-\omega/\Omega_c}\Theta(\omega-\omega_0)\,.
\end{equation}
Similarly, the bath correlator reads
\begin{multline*}
  K_\mathrm{wg}({\mathbf r},t)=\\
 \int_0^\infty d\omega\,J_\mathrm{wg}(\omega)
    \coth\left(\frac{\omega}{2T}\right)\cos(\omega t) \cos(r
    \sqrt{\omega^2 - \omega_0^2})\,.
\end{multline*}

\begin{figure}
  \centering
  \includegraphics{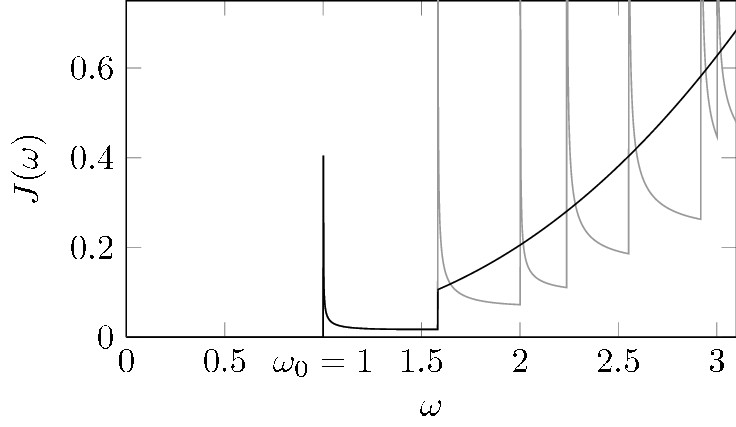}
  \caption{Coupling spectral density $J(\omega)$ of a quasi
    one-dimensional bath. We only consider the first transversal
    excitation and approximate the rest of the spectrum with the free
    bath (black line). The full spectrum is shown in gray.}
  \label{fig:tube-1}
\end{figure}

In order to approximate the full spectrum, we will use the approach
depicted in Fig.~\ref{fig:tube-1}: Only the first excitation is
considered and the rest of the spectrum is replaced with the free
bath. This is a good approximation as long as the oscillator frequency
$\Omega_0$ is close to the frequency of the first excitation
$\omega_0$.

\section{Numerics}
\label{sec5}

Due to the inverse square root divergence of the coupling spectral
density, the required numerical effort is larger than in the free
case. We were able to achieve good convergence by using the non-uniform
distribution of grid points,
\begin{eqnarray*}
s_k=\frac{s_\mathrm{max}-\omega_0}{(n_\mathrm{grid})^2}k^2+\omega_0
\end{eqnarray*}
resulting in the $k$-dependent spacing
\begin{eqnarray*}
\triangle s_k=\frac{2k}{(n_\mathrm{grid})^2}(s_\mathrm{max}-\omega_0)\,.
\end{eqnarray*}

\begin{figure}
  \centering
  \includegraphics{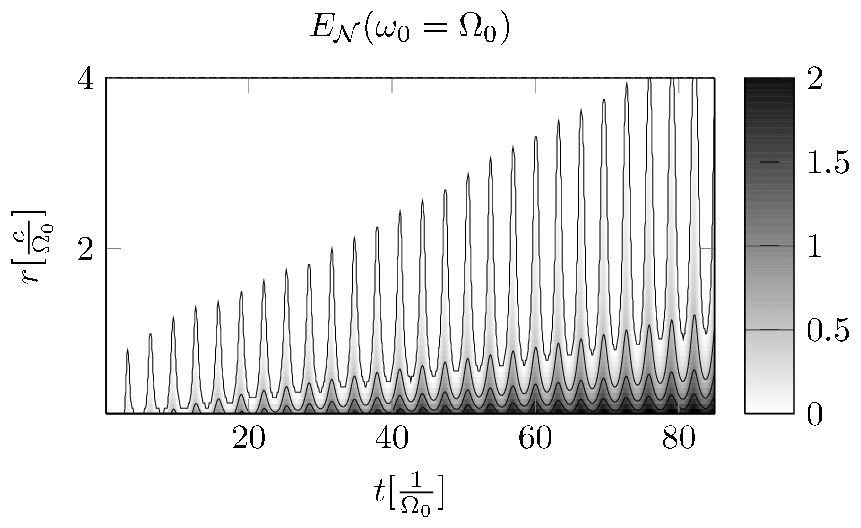}
  \caption{Dependence of the time evolution of the entanglement \EN{}
    on the spatial separation $r$ if the systems is placed inside of a
    waveguide. The value of the entanglement is encoded in the shade
    of gray as indicated by the scale on the right side. Additionally
    contour lines are drawn at the values $\EN = 0$, $0.5$, $1$, and
    $1.5$. The parameters are $\Omega_c = 10\,\Omega_0$, $\gamma =
    0.05\,\Omega_0$, and $T = 0\,\Omega_0$ with an initially
    separable single mode squeezed state $\kappa = 10$.}
  \label{fig:tube-2}
\end{figure}

The results of the numerical calculation are shown in the form of a
density plot in Fig.~\ref{fig:tube-2}. Shades of gray encode the value
of the entanglement which is drawn as function of time (x-axis) and
separation (y-axis).

\begin{figure}
  \centering
  \includegraphics{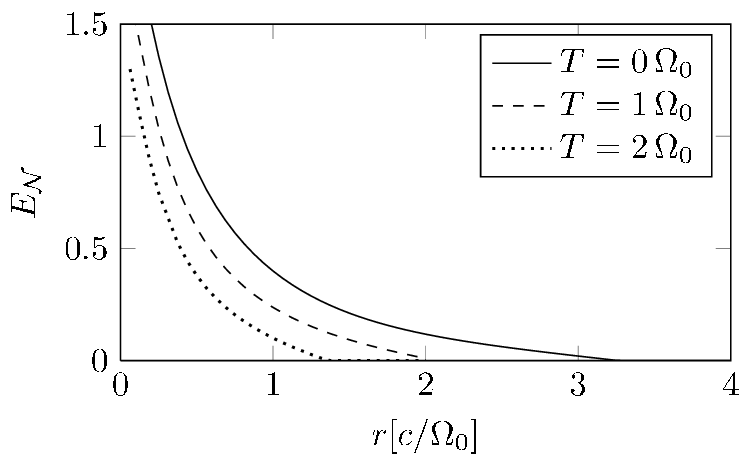}
  \caption{Distance dependence of \EN{} at $t=60/\Omega_0$ for
    different temperatures. All other parameters are the same as
    in Fig.~\ref{fig:tube-2}}
  \label{fig:tube-3}
\end{figure}

We again start with a separable state, which now is a single mode
squeezed state with the squeezing parameter $\kappa = 10$
\begin{equation}\label{eq:initial}
  \operatorname{Cov}(0) =
  \operatorname{diag}(1/(\kappa\Omega_0), 1/(\kappa\Omega_0),
  \kappa\Omega_0, \kappa\Omega_0)\,.
\end{equation}
The ground state corresponds to $\kappa = 1$. We note that
increasing the squeezing increases the absolute value of the
entanglement but does not change the distance over which it is
generated. The other parameters are comparable to the free case: The
coupling strength is set to $\gamma = 0.05\,\Omega_0$ and the cut-off
frequency is $\Omega_c = 10\,\Omega_0$. This time, however,
entanglement is generated over distances which are more than two
orders of magnitude larger than in the free bath. The effect is most
pronounced if the frequency of the peak $\omega_0$ is close to the
frequency of the oscillators $\Omega_0$.

Fig.~\ref{fig:tube-3} shows the exact distance dependence. The plot
corresponds to a vertical cut through Fig.~\ref{fig:tube-2} at time $t
= 60/\Omega_0$. Increasing the temperatures decreases the entanglement
generation, but the effect persists even for comparably large
temperatures of order $\Omega_0$.

\section{Analytical Model}\label{sec6}

\begin{figure}[h]
\includegraphics{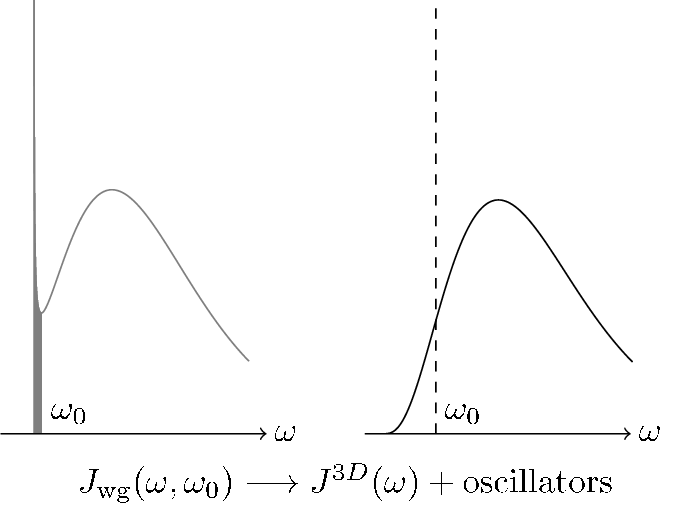}
\caption{Schematical illustration of the replacement of the van Hove coupling spectral density by the 
free coupling spectral density and effective oscillators with frequency $\omega_0$.}\label{vanhove}
\end{figure}

As we have seen in section (\ref{sec5}), boundary conditions 
which are imposed on the bath modes have an impact 
on the generation of entanglement.
According to equation (\ref{eq:vanHovespec}), the spectral weight of the frequencies in an
interval $\delta \omega$ centered around
the van Hove peak is significantly larger than the spectral weight of frequencies
in an interval $\delta \omega$  which is distant from the singularity.
This fact can be used for a derivation
of approximate analytical solutions of the model discussed in the previous section.
Therefore, we take into account only the first van Hove singularity at $\omega_0=\sqrt{8}\pi/a$
and replace the coupling spectral density (\ref{eq:vanHovespec}) by two effective macroscopic oscillators, 
representing the coherently oscillating symmetric and antisymmetric field modes around the van Hove singularity.
Furthermore, we take into account for the incoherently oscillating modes
of the bath by a background spectral density, given by (\ref{eq:specdens}).
This replacement, schematically depicted in fig. \ref{vanhove},
will be a suitable approximation as long as the modes within the van Hove peak are oscillating 
coherently, that is, for $t,r<1/\delta \omega$. 
The effective oscillator representing the symmetric modes will determine strongly the 
generation of entanglement whereas dissipation
and decoherence is determined by $J^{3D}$.

We choose $\mathbf{r}=r\mathbf{e}_1$ and decompose the system variables and the field
modes in symmetric and antisymmetric compositions with respect
to the $\mathbf{e}_1$-direction, that is
\begin{eqnarray*}
\hat{\phi}_{\mathbf{k},S/A}&=&\frac{1}{\sqrt{2}}\left(\hat{\phi}_{k_1,k_2,k_3}\pm\hat{\phi}_{-k_1,k_2,k_3}\right)\\
\hat{\Pi}_{\mathbf{k},S/A}&=&\frac{1}{\sqrt{2}}\left(\hat{\Pi}_{k_1,k_2,k_3}\pm\hat{\Pi}_{-k_1,k_2,k_3}\right)\\
\hat{Q}_{S/A}&=&\frac{1}{\sqrt{2}}(\hat{Q}_1\pm \hat{Q}_2)\\
\hat{P}_{S/A}&=&\frac{1}{\sqrt{2}}(\hat{P}_1\pm \hat{P}_2)\,.
\end{eqnarray*}
We find that the effective coupling oscillator interacts only with the symmetric mode $\hat{Q}_S$. 
Within this approximation, the Hamiltonian (\ref{Hamilt}) reads
\begin{eqnarray*}
\hat{H}=\hat{H}_{\mathrm{B},S}+\hat{H}_{\mathrm{B},A}+\hat{H}_{0,S}+\hat{H}_{0,A}+\hat{H}_\mathrm{int}
\end{eqnarray*}
with
\begin{eqnarray*}
\hat{H}_{\mathrm{B},S/A}&=&\frac{1}{2}\sum_{k_1>0,k_2,k_3}\left(\hat{\Pi}_{\mathbf{k},S}\hat{\Pi}_{-\mathbf{k},S/A}+
k^2 \hat{\phi}_{\mathbf{k},S}\hat{\phi}_{-\mathbf{k},S/A}\right)\,,\\
\hat{H}_{0,S}&=&\frac{1}{2}\left(\hat{P}_S^2+\Omega_S^2\hat{Q}_S^2\right)+\frac{1}{2}\left(\hat{p}_S^2+\omega_0^2\hat{q}^2_S\right)- g \,\hat{p}_S\,\hat{Q}_S\,,\\
\hat{H}_{0,A}&=&\frac{1}{2}\left(\hat{P}_A^2+\Omega_A^2\hat{Q}_A^2\right)+\frac{1}{2}\left(\hat{p}_A^2+\omega_0^2\hat{q}_A^2\right)\,,\\
\hat{H}_\mathrm{int}&=&-\sum_{k_1>0,k_2,k_3}2g_k
\bigg[\hat{\Pi}_{\mathbf{k},S}\hat{Q}_S\cos\left(\frac{k_1r}{2}\right)\\
& &\hspace{2cm}+i\hat{\Pi}_{\mathbf{k},A}\hat{Q}_A\sin\left(\frac{k_1r}{2}\right)\bigg]\,.
\end{eqnarray*}
Here we denoted the canonical position and momentum variables of 
the effective oscillators with $\hat{q}_{S/A}$ and $\hat{p}_{S/A}$, respectively.
The coupling constant $g$ depends on the coupling spectral density and will be determined below.
Furthermore, the counterterms of the Hamiltonian lead to a renormalization of the system oscillator 
modes, that is
\begin{eqnarray*}
\Omega_{S/A}&=&\sqrt{\Omega_0^2+\int_0^\infty d\omega \frac{J^{3D}(\omega)}{\omega}\left(1\pm\frac{\sin(\omega r)}{\omega r}\right)}\,.
\end{eqnarray*}
The Hamiltonian $\hat{H}_\mathrm{S}$ can be diagonalized by means of the 
transformations
\begin{eqnarray}\label{trafo}
\xi&=&\frac{\omega_0^2-\Omega_S^2-g^2+\sqrt{(\Omega_S^2+g^2-\omega_0^2)^2+4 g^2\omega_0^2}}{2g\omega_0}\,,\nonumber\\
\hat{\bar{Q}}_1&=&\frac{\omega_0 \hat{Q}_S-\xi \hat{p}}{\omega_0 \sqrt{1+\xi^2}}\,,\quad \hat{\bar{Q}}_2=\frac{\omega_0 \hat{q}-\xi \hat{P}_S}{\Omega_2\sqrt{1+\xi^2}}\,,\nonumber\\
\hat{\bar{P}}_1&=&\frac{\hat{P}_S+\xi \omega_0 \hat{q}}{\sqrt{1+\xi^2}}\,,\quad \hat{\bar{P}}_2=\frac{\Omega_2 (\hat{p}+\xi\omega_0 \hat{Q}_S)}{\omega_0\sqrt{1+\xi^2}}\,,
\end{eqnarray}
and adopts the canonical form
\begin{eqnarray*}
\hat{H}_\mathrm{S}&=&\frac{1}{2}\sum_{i=1,2}(\hat{\bar{P}}_i^2+\Omega_i^2 \hat{\bar{Q}}_i^2)\nonumber\\
& +&\frac{1}{2}(\hat{\bar{P}}_A^2+
\Omega_A^2 \hat{\bar{Q}}_A^2+\hat{\bar{p}}_A^2+\omega_0^2 \hat{\bar{q}}_A^2))\,.
\end{eqnarray*}
The eigenmodes are given by
\begin{eqnarray*}
\Omega_1&=&\sqrt{\frac{1}{2}\left(\Omega_S^2+g^2+\omega_0^2+\sqrt{(\Omega_S^2+g^2-\omega_0^2)^2+4 g^2\omega_0^2}\right)}\nonumber\\
\Omega_2&=&\sqrt{\frac{1}{2}\left(\Omega_S^2+g^2+\omega_0^2-\sqrt{(\Omega_S^2+g^2-\omega_0^2)^2+4 g^2\omega_0^2}\right)}\nonumber\,.
\end{eqnarray*}
The time-evolution of the system oscillators and the coupling
oscillator will be treated exactly, for the incoherently
oscillating modes we will use a master equation approach. 

The time evolution of the density matrix is determined by the differential equation 
$\hat{\rho}_\mathrm{SB}=-i[\hat{H},\hat{\rho}_\mathrm{SB}]$.
Assuming that the bath correlators decay on a time scale which is much shorter
than the time scale of the system oscillators allows for a Born-Markov-approximation \cite{S07}.
The density matrix of the system factorizes in a symmetric and an antisymmetric part, $\hat{\rho}_\mathrm{S}=\hat{\rho}_S\otimes\hat{\rho}_A$,
which obey the differential equations
\begin{eqnarray}\label{rhoS}
\partial_t\hat{\rho}_{S/A}&=&-i\sum_{i=1,2}\left[\hat{H}_{S/A},\hat{\rho}_S\right]\\
& &-\int_0^\infty dt\,\nu_{S/A}(t,r)
[\hat{Q}_{S/A},[\hat{Q}_{S/A}(-t),\hat{\rho}_{S/A}]]\nonumber\\
& &+i\int_0^\infty dt\,\mu_{S/A}(t,r)
[\hat{Q}_{S/A},\{\hat{Q}_{S/A}(-t),\hat{\rho}_{S/A}\}]\nonumber\,.
\end{eqnarray}
The bath correlators $\nu_{S/A}$ and $\mu_{S/A}$ are given by
\begin{eqnarray*}
\nu_{S/A}(t,r)& &\\
& &\hspace{-1.5cm}=\frac{1}{2}\int d\omega J^{3D}(\omega)\coth\left(\frac{\omega}{2T}\right)\left(1\pm\frac{\sin(\omega r)}{\omega r}\right)\cos(\omega t)\,,\nonumber\\
\mu_{S/A}(t,r)&=&\frac{1}{2}\int d\omega J^{3D}(\omega)\left(1\pm\frac{\sin(\omega r)}{\omega r}\right)\sin(\omega t)\,.
\end{eqnarray*}
Decoherence and dissipation is completely determined by the double commutators involving
eight different system-bath-correlators for the symmetric and antisymmetric bath
modes, respectively.
The decoherence rate is determined by system-bath-correlators $\alpha_{1,S/A}(\Omega)=\int_0^\infty dt\,\nu_{S/A}\cos(\Omega t)$,
the correlator $\alpha_{2,S/A}(\Omega)=-\int_0^\infty dt\, \nu_{S/A}\sin(\Omega t)/\Omega$ is called anomalous-diffusion coefficients,
$\alpha_{3,S/A}(\Omega)=-\int_0^\infty dt\,\mu_{S/A}\cos(\Omega t)$ introduces a the lamb shift and finally $\alpha_{4,S/A}(\Omega)
=\int_0^\infty dt\, \mu_{S/A}\sin(\Omega t)/\Omega$
determines the strength of dissipation.
The frequency $\Omega$ is equal to $\bar{\Omega}_{1}$ or $\bar{\Omega}_{2}$ for the correlators with index $S$
and equal to $\omega$ or $\Omega_{A}$ for the correlators with index $A$.
Explicit expressions are given in the Appendix, see equations (\ref{corr1})-(\ref{corr4}).

\subsection{Induced oscillator coupling}

The effective coupling constant $g$ can be roughly estimated from the expressions
of the correlators $\nu_{S}$ or $\mu_{S}$.
Splitting the wave vector as in section (\ref{sec4}) and replacing the $n_1$-summation by an integral
we find
\begin{eqnarray*}
 \nu_S(t,r)&=&\int_0^\infty d\omega\sum_{n_2,n_3=1}^{\frac{\omega_0^2}{2}(n_2^2+n_3^2)<\omega^2} \frac{\omega_0^2 J^{3D}(\omega)}
{\pi \omega\sqrt{\omega^2-\frac{\omega_0^2}{2}(n_2^2+n_3^2)}}\nonumber\\
& &\hspace{-1cm}\times\coth\left(\frac{\omega}{2T}\right)\cos(\omega t)
\cos^2\left(\frac{r\sqrt{\omega^2-\frac{\omega_0^2}{2}(n_1^2+n_2^2)}}{2}\right)\,.\end{eqnarray*}
Restricting ourselves to the first van Hove singularity, $n_2=n_2=1$, and integrating
from $\omega_0$ to $\omega_0+\delta\omega$, we deduce the relation
\begin{eqnarray*}
g^2\langle \hat{p}(t)\hat{p}\hat{\rho}_B\rangle&=&g^2\frac{\omega_0}{2}
\cos(\omega_0 t)
\coth\left(\frac{\omega_0}{2T}\right)\\
&\approx&\frac{\omega_0}{\pi}J^{3D}(\omega_0)\sqrt{\frac{2\delta\omega}{\omega_0}}\cos(\omega_0 t)
\coth\left(\frac{\omega_0}{2T}\right)\nonumber\,.
\end{eqnarray*}
The modes within the van Hove peak are oscillating coherently
for distances $r\lesssim c/\delta\omega$ which introduces an effective
distance dependence of $g$.
Since this coupling constant would increase indefinitely 
for $r\rightarrow 0$, we need a natural cutoff for the smallest
distance possible which is given by $1/\Omega_c$.
Therefore we have $\delta\omega=1/(r+1/\Omega_c)$
and
\begin{eqnarray}\label{effcoupl}
g=g(r)\approx\sqrt{\frac{2J^{3D}(\omega_0)}{\pi}\sqrt{\frac{2}{\omega_0 r+\frac{\omega_0}{\Omega_c}}}}\,.
\end{eqnarray}

\subsection{Analytical solutions}

Although we restrict ourselves to Gaussian density matrices,
equations (\ref{rhoS}) lead to a system of coupled
nonlinear differential equations in the position representation.
For this reason we transform the density matrix into the ``$k-\Delta$''-representation \cite{UZ89}.
Since the procedure is the same for $\hat{\rho}_S$ and $\hat{\rho}_A$, we restrict ourselves
to the former one.
The transformed density matrix is given by
\begin{eqnarray*}
\tilde{\rho}(\mathbf{k},\mathbf{\Delta})=\mathrm{Tr}\left(\hat{\rho}_S e^{i\left(\mathbf{k} \mathbf{\hat{\bar{Q}}}+
\mathbf{\Delta}\mathbf{\hat{\bar{P}}}\right)}\right)\,.
\end{eqnarray*}
with the vectors $\mathbf{k}=(k_1,k_2)^T$, $\mathbf{\Delta}=(\Delta_1,\Delta_2)^T$,
$\mathbf{\hat{\bar{Q}}}=(\hat{\bar{Q}}_1,\hat{\bar{Q}}_2)$ and $\mathbf{\hat{\bar{P}}}=(\hat{\bar{P}}_1,\hat{\bar{P}}_2)$.
This representation is related to the Wigner-Distribution via a double Fourier transformation
and has in position representation the form
\begin{eqnarray*}
\tilde{\rho}(\mathbf{k},\mathbf{\Delta})=\int dx_1dx_2 e^{i\mathbf{k}\mathbf{x}}\rho_S\left(\mathbf{x}+\frac{\mathbf{\Delta}}{2},\mathbf{x}-\frac{\mathbf{\Delta}}{2}\right)
\end{eqnarray*}
where $\mathbf{x}=(x_1,x_2)$ labels the diagonal elements of the density matrix.
Using this particular representation we find that the master equation is linear
in the derivatives with respect to $k_i$ and $\Delta_i$.
With an Gaussian ansatz for the density matrix $\rho(\mathbf{k},\mathbf{\Delta})$ we end up with a linear first order system
of differential equations, see equation (\ref{DGLs}) in the Appendix.

The dissipative part of equation (\ref{rhoS}) leads to a coupling among 14 differential 
equations, thus the eigenmodes cannot be found analytically. 
However, neglecting terms $\mathcal{O}(\gamma\xi)$ in the \textit{dissipative} part of equation (\ref{DGLPrimed})
allows for an analytical treatment of (\ref{DGLs}).
After removing the higher order couplings we are left with the usual 4 double commutators
known from the Caldeira-Leggett model (see e.g.\cite{S07}).

The expectation values of the anticommutators, and therefore the negativity, can be expressed
in terms of the solutions $c_i(t)$ of the differential equations (see equations \ref{comm}).
From the approximate solutions (\ref{Approx1}) - (\ref{Approx3})
we find that the damping of the eigenmodes is completely determined
by the dissipation-correlator.
The lambshift modifies the eigenmodes $\bar{\Omega}_{1/2}$.
We found that analytical expressions coincide very well with the numerical
integration of the differential equations (\ref{DGLs}) for small and intermediate times.
\begin{figure}[h]
\includegraphics{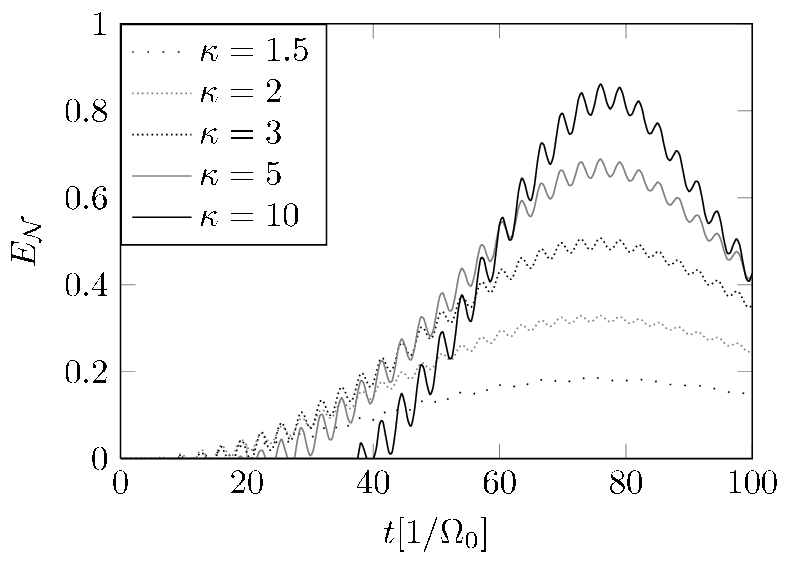}
\caption{Dependence of the time evolution of the entanglement on different squeezing parameters.
The remaining parameters are $s=3$, $\gamma=0.05\Omega_0$, $\Omega_c=3\Omega_0$ and $r=2[c/\Omega_0]$.
}\label{kappa}
\end{figure}
\begin{figure}[h]
\includegraphics{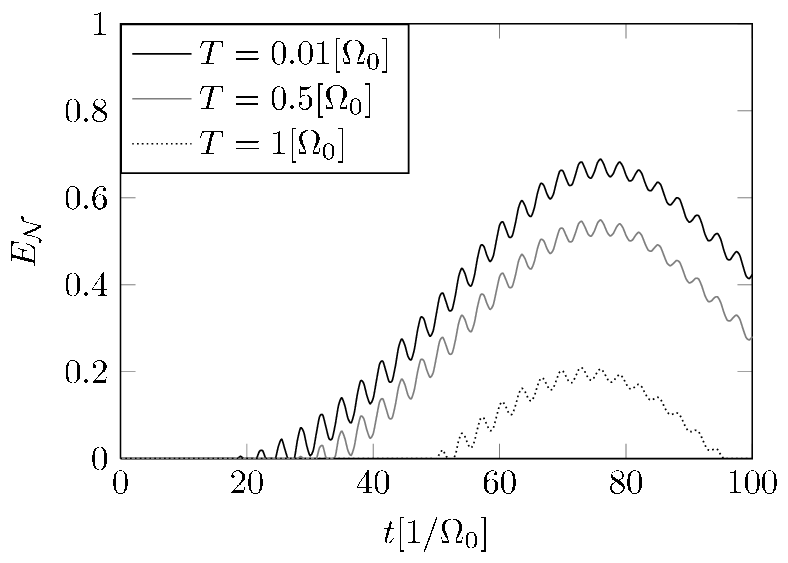}
\caption{Dependence of the time evolution of the entanglement on different temperatures.
The squeezing parameter is $\kappa=5$ and the remaining parameters are the same as in Fig.~\ref{kappa}.}\label{temp}
\end{figure}

\subsection{Results}

We consider the system oscillators to be initially in a Gaussian state with squeezing
parameter $\kappa$, see equation (\ref{eq:initial}), the 
effective coupling oscillator is assumed to be in the ground state.
From Fig.~\ref{kappa} we deduce that the maximum value of the 
negativity depends crucially on the initial squeezing.
Due to the dissipative time evolution, the system oscillators
relax within intermediate times from the squeezed and nonentangled state to an entangled state with lower energy. 
In contrast, preparing the system initially in the ground state, that is $\kappa=1$, only negativity of order $\gamma^2/\Omega_0^2$ is generated.

The temperature dependence is very weak for $T\ll \Omega_0$ whereas
for $T=\mathcal{O}(\Omega_0)$, the negativity decreases significantly, see Fig.~\ref{temp}.

\begin{figure}[t]
\includegraphics{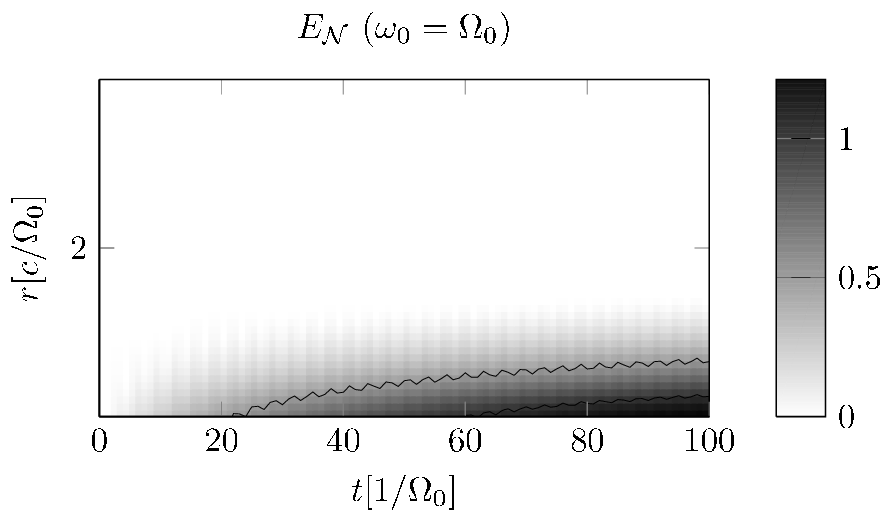}
\caption{Dependence of the time evolution of the entanglement on the spatial separation
of the system oscillators without coupling oscillator.
The parameters are $\kappa=5$, $s=3$, $\gamma=0.05\Omega_0$, $\Omega_c=3\Omega_0$ and $T=0.01[\Omega_0]$.}\label{gzero}
\end{figure}

\begin{figure}[t]
\includegraphics{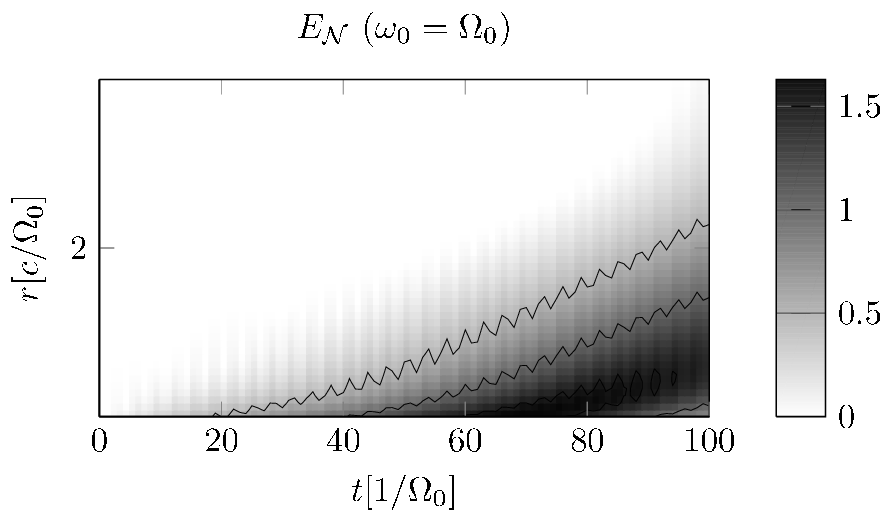}
\caption{Dependence of the time evolution of the entanglement on the spatial separation
of the system oscillators.
The parameters are $\kappa=5$, $s=3$, $\gamma=0.05\Omega_0$, $\Omega_c=3\Omega_0$ and $T=0.01[\Omega_0]$.}\label{distance}
\end{figure}

In virtue of the van Hove singularity we find significant entanglement
over distances $r\gg\mathcal{O}(c/\Omega_c)$ as can be deduced
from Fig.~\ref{distance}, whereas for $g=0$, which corresponds 
to the time evolution of the free space environment, we find no entanglement for distances that are larger than $\mathcal{O}(c/\Omega_c)$, see Fig.~\ref{gzero}.
From the expressions (\ref{corr1}) and (\ref{corr2}) we see that the generation of entanglement for $g=0$ is related to the
difference between the decoherence and dissipation rates of the symmetric and antisymmetric mode, respectively.

Although various approximations were necessary in order to derive
analytical expressions for the time evolution, we find good qualitative
agreements with the exact numerical calculations, compare Fig.~\ref{distance}
with Fig.~\ref{fig:tube-2}.


\begin{figure}[ht]
\includegraphics{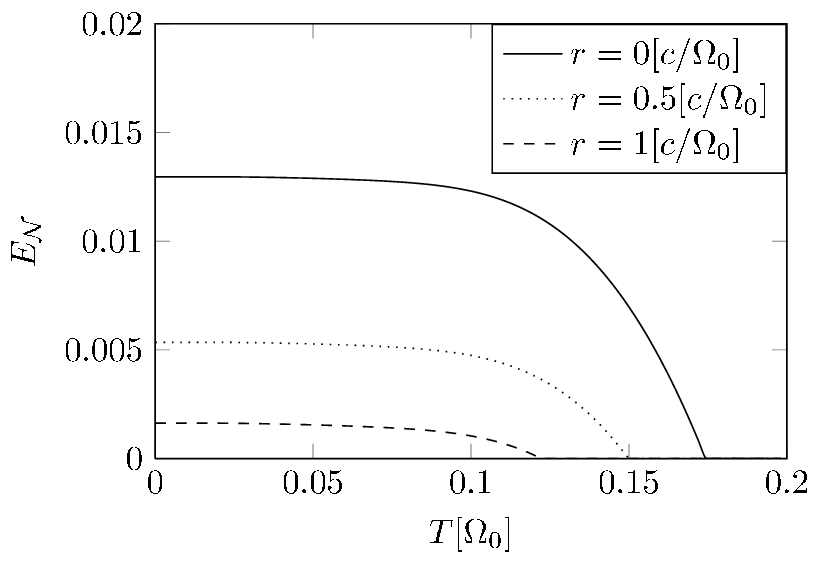}
\caption{Dependence of the asymptotic entanglement on the temperature for different spatial separation
of the system oscillators.
The parameters are $s=1$, $\gamma=0.01\Omega_0$, $\Omega_c=3\Omega_0$ and $g=1[\Omega_0]$.}\label{asymtemp}
\end{figure}

\begin{figure}[ht]
\includegraphics{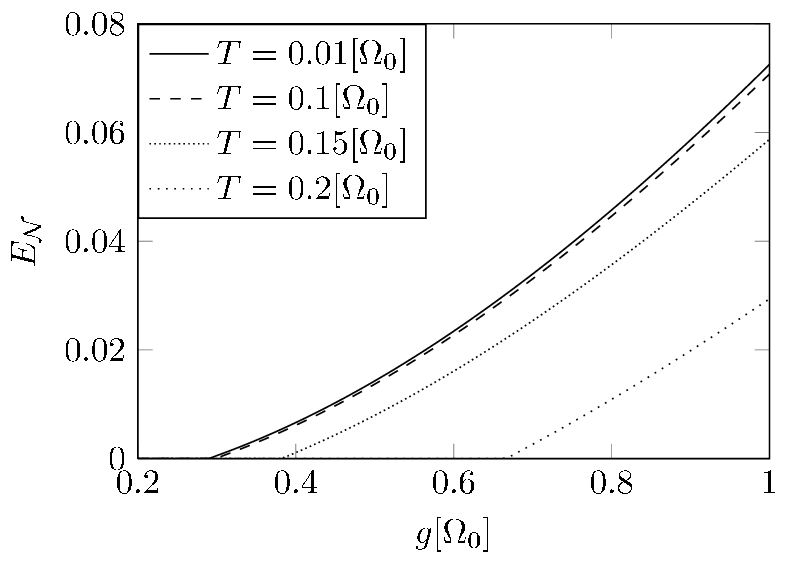}
\caption{Dependence of the asymptotic entanglement on the coupling $g$ for different temperatures.
The parameters are $s~=~1$, $\gamma=0.01\Omega_0$, $\Omega_c=3\Omega_0$
and $r~=~1[c/\Omega_0] $.}\label{asymg}
\end{figure}

Taken the approximate relation (\ref{effcoupl}) for granted, all entanglement vanishes in the asymptotic limit.
However, treating $g$ and $\gamma$ independently from each other we find for sufficiently large $g$ nonzero asymptotic negativity
which can also be determined from the thermal expectation value of the covariance matrix up to corrections of order $\mathcal{\gamma\xi}$.
In Fig.~\ref{asymtemp}, we depict the temperature-dependence of the asymptotic entanglement for $g=\Omega_0$
for different values of $r$.
From Fig.~\ref{asymg} we deduce that a critical minimal value of $g$ is necessary in order to
observe asymptotically a nonzero negativity.

\section{Conclusions}

We investigated the generation of entanglement
of remote quantum systems via a bosonic heat bath.
Starting from an Lagrangian formulation describing
the coupling of two remote harmonic oscillators
to a scalar field environment, we derived the Langevin
equations of motion.
The coupling of the oscillators to the field momenta
guaranteed the positivity of the corresponding Hamiltonian.
This led to an counter-term which was already proposed in \cite{ZQK09} 
in order to preserve causality and to avoid runaway solutions.

In case of a free space environment the generated entanglement
drops quickly to zero if the distance between the 
ocsillators exceeds the spatial extension 
of the individual quantum systems.
For one and three spatial dimensions, we found that 
this statement is independent of the initial conditions.

In contrary, by imposing boundary conditions on the heat bath
it is possible to generate entanglement over distances
which exceeds significantly the spatial extension of the quantum systems.
In particular, we considered a waveguide with a transversal 
extension corresponding to the inverse system frequency.
The boundary condition leads to a van Hove peak in the coupling spectral
density.
The entanglement can be enhanced significantly if the van Hove peak
is in resonance with the remote quantum systems.
Furthermore, the situation can be improved
if the quantum systems are prepared initially in a 
strongly squeezed state.

Considering the coherently oscillating modes as an effective oscillator coupling
the quantum systems to each other, we derived analytical solutions
which are qualitatively in good agreement to the numerical results.

\section{Appendix}
\label{sec8}
\subsection{Differential equations}
In order to avoid a cluttering of indices we will restrict ourselves to the 
density matrix for the symmetric mode and suppress the label $S$.
The equations determining the antisymmetric mode follow from a replacement
of the eigenmodes and canonical variables in the limit $\xi\rightarrow0$.
The time evolution for the oscillator variables is given by
\begin{eqnarray}
\hat{\bar{Q}}_{1/2}(t)=\hat{\bar{Q}}_{1/2} \cos(\bar{\Omega}_{1/2}t)+\frac{1}{\bar{\Omega}_{1/2}}\hat{\bar{P}}_{1/2}\sin(\bar{\Omega}_{1/2}t)\nonumber\,,
\end{eqnarray}
whereas for momentum variables we find
\begin{eqnarray}
\bar{P}_{1/2}(t)=\bar{P}_{1/2} \cos(\bar{\Omega}_{1/2}t)-\bar{\Omega}_{1/2}\bar{Q}_{1/2}\sin(\bar{\Omega}_{1/2}t)\nonumber
\end{eqnarray}
Using relations (\ref{trafo}) we can decompose the differential equation (\ref{rhoS}) with respect to 
the (time-independent) operators
$\bar{Q}_{1/2}$ and $\bar{P}_{1/2}$, that is
\begin{eqnarray}
\dot{\hat{\rho}}&=&-i\sum_{i=1,2}\frac{1}{2}\left[\hat{\bar{P}}_i^2+\bar{\Omega}_i^2\hat{\bar{Q}}_i^2,\hat{\rho}\right]\label{DGLPrimed}\\
& &-\frac{\alpha_1(\bar{\Omega}_1)}{1+\xi^2}\left([\hat{\bar{Q}}_1,[\hat{\bar{Q}}_1,\hat{\rho}]]+\frac{\xi}{\bar{\Omega}_2}[\hat{\bar{P}}_2,[\hat{\bar{Q}}_1,\hat{\rho}]]\right)\nonumber\\
& &-\frac{\alpha_2(\bar{\Omega}_1)}{1+\xi^2}\left([\hat{\bar{Q}}_1,[\hat{\bar{P}}_1,\hat{\rho}]]+\frac{\xi}{\bar{\Omega}_2}[\hat{\bar{P}}_2,[\hat{\bar{P}}_1,\hat{\rho}]]\right)\nonumber\\
& &-\frac{\xi\,\alpha_1(\bar{\Omega}_2)}{(1+\xi^2) \bar{\Omega}_2^2}\bigg(\xi[\hat{\bar{P}}_2,[\hat{\bar{P}}_2,\hat{\rho}]]+\bar{\Omega}_2[\hat{\bar{Q}}_1,[\hat{\bar{P}}_2,\hat{\rho}]]\bigg)\nonumber\\
& &+\frac{\xi\,\alpha_2(\bar{\Omega}_2)}{1+\xi^2}\bigg(\xi[\hat{\bar{P}}_2,[\hat{\bar{Q}}_2,\hat{\rho}]] +\bar{\Omega}_2[\hat{\bar{Q}}_1,[\hat{\bar{Q}}_2,\hat{\rho}]]\bigg)\nonumber\\
& &-i\frac{\alpha_3(\bar{\Omega}_1)}{1+\xi^2}\left([\hat{\bar{Q}}_1,\{\hat{\bar{Q}}_1,\hat{\rho}\}]+ \frac{\xi}{\bar{\Omega}_2}[\hat{\bar{P}}_2,\{\hat{\bar{Q}}_1,\hat{\rho}\}]\right)\nonumber\\
& &-i\frac{\alpha_4(\bar{\Omega}_1)}{1+\xi^2}\left([\hat{\bar{Q}}_1,\{\hat{\bar{P}}_1,\hat{\rho}\}]+ \frac{\xi}{\bar{\Omega}_2}[\hat{\bar{P}}_2,\{\hat{\bar{P}}_1,\hat{\rho}\}]\right)\nonumber\\
& &-i\frac{\xi\,\alpha_3(\bar{\Omega}_2)}{(1+\xi^2)\bar{\Omega}_2^2}\bigg(\xi[\hat{\bar{P}}_2,\{\hat{\bar{P}}_2,\hat{\rho}\}] +\bar{\Omega}_2[\hat{\bar{Q}}_1,\{\hat{\bar{P}}_2,\hat{\rho}\}]\bigg)\nonumber\\
& &+ i\frac{\xi\,\alpha_4(\bar{\Omega}_2)}{1+\xi^2}\bigg(\xi[\hat{\bar{P}}_2,\{\hat{\bar{Q}}_2,\hat{\rho}\}]+\bar{\Omega}_2[\hat{\bar{Q}}_1,\{\hat{\bar{Q}}_2,\hat{\rho}\}]\bigg)\nonumber\,,
\end{eqnarray}
where $\xi$ was defined in equation (\ref{trafo}) and the $\alpha_i$ are 
correlation functions that are given explicitly in section \ref{bathcorr}.
The equations of motion in the ``$k-\Delta$''-representation can be constructed by wrapping
\begin{eqnarray}
\mathrm{Tr}\left(\hat{D}...\right)= \mathrm{tr}\left(e^{i\left(\mathbf{k} \mathbf{\hat{\bar{Q}}}+
\mathbf{\Delta}\mathbf{\hat{\bar{P}}}\right)}...\right)
\end{eqnarray}
over equation (\ref{DGLKDELTA}).
Using the relations
\begin{eqnarray}
\mathrm{Tr} (\hat{D}[\hat{\bar{Q}}_i,[\hat{\bar{Q}}_j,\hat{\rho}]])&=&\Delta_i\Delta_j\tilde{\rho}\nonumber\\
\mathrm{Tr} (\hat{D}[\hat{\bar{Q}}_i,[\hat{\bar{P}}_j,\hat{\rho}]])&=&-\Delta_ik_j\tilde{\rho}\nonumber\\
\mathrm{Tr}(\hat{D}[\hat{\bar{P}}_i,[\hat{\bar{Q}}_j,\hat{\rho}]])&=&-k_i\Delta_j\tilde{\rho}\nonumber\\
\mathrm{Tr} (\hat{D}[\hat{\bar{P}}_i,[\hat{\bar{P}}_j,\hat{\rho}]])&=&k_i k_j\tilde{\rho}\nonumber\\
\mathrm{Tr} (\hat{D}[\hat{\bar{Q}}_i,\{\hat{\bar{Q}}_j,\hat{\rho}\}])&=&-2 i\Delta_i\partial_{k_j}\tilde{\rho}\nonumber\\
\mathrm{Tr} (\hat{D}[\hat{\bar{Q}}_i,\{\hat{\bar{P}}_j,\hat{\rho}\}])&=&-2 i\Delta_i\partial_{\Delta_j}\tilde{\rho}\nonumber\\
\mathrm{Tr} (\hat{D}[\hat{\bar{P}}_i,\{\hat{\bar{Q}}_j,\hat{\rho}\}])&=&2i k_i\partial_{k_j}\tilde{\rho}\nonumber\\
\mathrm{Tr} (\hat{D}[\hat{\bar{P}}_i,\{\hat{\bar{P}}_j,\hat{\rho}\}])&=&2i k_i\partial_{\Delta_j}\tilde{\rho}\label{A11}\,,
\end{eqnarray}
we arrive at
\begin{eqnarray}
\dot{\tilde{\rho}}&=&\left(k_1-\frac{2}{1+\xi^2}\alpha_4(\bar{\Omega}_1)\left(\Delta_1-\frac{\xi}{\bar{\Omega}_2}k_2\right)\right)\partial_{\Delta_1}\tilde{\rho}\nonumber\\
& &+\left(k_2-\frac{2\xi}{(1+\xi^2)\bar{\Omega}_2^2}\alpha_3(\bar{\Omega}_2)( \bar{\Omega}_2\Delta_1-\xi k_2)\right)\partial_{\Delta_2}\tilde{\rho}\nonumber\\
& &-\left(\bar{\Omega}_1^2\Delta_1+\frac{2}{1+\xi^2}\alpha_3(\bar{\Omega}_1)\left(\Delta_1-\frac{\xi k_2}{\bar{\Omega}_2}\right)\right)\partial_{k_1}\tilde{\rho}\nonumber\\
& &-\left(\bar{\Omega}_2^2 \Delta_2-\frac{2\xi}{1+\xi^2}\alpha_4(\bar{\Omega}_2)(\bar{\Omega}_2\Delta_1-\xi k_2)\right)\partial_{k_2}\tilde{\rho}\nonumber\\
& &+\frac{1}{1+\xi^2}\bigg[-\alpha_1(\bar{\Omega}_1)\left(\Delta_1^2-\frac{\xi}{\bar{\Omega}_2}k_2\Delta_1\right)\nonumber\\
& &\hspace{1.5cm}+\alpha_2(\bar{\Omega}_1)\left(k_1\Delta_1-\frac{\xi}{\bar{\Omega}_2}k_1k_2\right)\nonumber\\
& &\hspace{1.5cm}+\frac{\xi}{\bar{\Omega}_2^2}\alpha_1(\bar{\Omega}_2)(\bar{\Omega}_2 k_2\Delta_1-\xi\,k_2^2)\nonumber\\
& &\hspace{1.5cm}+\xi\,\alpha_2(\bar{\Omega}_2)(\bar{\Omega}_2 \Delta_1\Delta_2-\xi\,k_2\Delta_2)\bigg]\tilde{\rho}\,.\label{DGLKDELTA}
\end{eqnarray}
With the Gaussian ansatz
\begin{eqnarray}
\tilde{\rho}_{Q_S,q}&=&\exp\big[-c_1k_1^2-c_2 k_1 \Delta_1-c_3\Delta_1^2-i c_4 k_1 -i c_5 \Delta_1\nonumber\\
& &-c_6 k_2^2-c_7 k_2 \Delta_2-c_8 \Delta_2^2
 - ic_9 k_2 -i c_{10} \Delta_2 \nonumber\\
& &-c_{11}k_1 k_2 -c_{12}k_1 \Delta_2-c_{13}k_2 \Delta_1 -c_{14}\Delta_1 \Delta_2\big]\label{A19}
\end{eqnarray}
we arrive at the first order system
\begin{eqnarray*}
\dot{c}_1&=&c_2\\ 
\dot{c}_2&=&-\left(2 \bar{\Omega}_1^2+\frac{4}{1+\xi^2}\alpha_3(\bar{\Omega}_1)\right)c_1+2c_3\nonumber\\
& &-\frac{1}{1+\xi^2}\bigg(2\alpha_4(\bar{\Omega}_1)c_2-2\xi\bar{\Omega}_2\alpha_4(\bar{\Omega}_2)c_{11}\\
& &+\frac{2\xi}{\bar{\Omega}_2}\alpha_3(\bar{\Omega}_2)c_{12}+\alpha_2(\bar{\Omega}_1)\bigg)\\
\dot{c}_3&=&-\left(\bar{\Omega}_1^2+\frac{2}{1+\xi^2}\alpha_3(\bar{\Omega}_1)\right)c_2\nonumber\\
& &-\frac{1}{1+\xi^2}\bigg(4\alpha_4(\bar{\Omega}_1)c_3-2\xi\bar{\Omega}_2\alpha_4(\bar{\Omega}_2)c_{13}\\
& &+\frac{2\xi}{\bar{\Omega}_2}\alpha_3(\bar{\Omega}_2)c_{14}-\alpha_1(\bar{\Omega}_1)\bigg)\\
\dot{c}_4&=&c_5\\
\dot{c}_5&=&-\left(\bar{\Omega}_1^2+\frac{2}{1+\xi^2}\alpha_3(\bar{\Omega}_1)\right)c_4\nonumber\\
& &-\frac{1}{1+\xi^2}\bigg(2\alpha_4(\bar{\Omega}_1)c_5-2\xi\bar{\Omega}_2\alpha_4(\bar{\Omega}_2)c_9\\
& &+\frac{2\xi}{\bar{\Omega}_2}\alpha_3(\bar{\Omega}_2)c_{10}\bigg)
\end{eqnarray*}
\begin{eqnarray*}
\dot{c}_6&=&-\frac{4 \xi^2}{1+\xi^2}\alpha_4(\bar{\Omega}_2)c_6+\left(1+\frac{2\xi^2}{(1+\xi^2)\bar{\Omega}_2^2}\alpha_3(\bar{\Omega}_2)\right)c_7\nonumber\\
& &+\frac{1}{1+\xi^2}\bigg(\frac{2\xi}{\bar{\Omega}_2}\alpha_3(\bar{\Omega}_1)c_{11}+\frac{2\xi}{\bar{\Omega}_2}\alpha_4(\bar{\Omega}_1)c_{13}\\
& & +\frac{\xi^2}{\bar{\Omega}_2^2}\alpha_1(\bar{\Omega}_2)\bigg)\\
\dot{c}_7&=&-2\bar{\Omega}_2^2 c_6-\frac{2\xi^2}{1+\xi^2}\alpha_4(\bar{\Omega}_2)c_7\\
& &+\left(2+\frac{4\xi^2}{(1+\xi^2)\bar{\Omega}_2^2}\alpha_3(\bar{\Omega}_2)\right)c_8\nonumber\\
& &+\frac{1}{1+\xi^2}\bigg(\frac{2\xi}{\bar{\Omega}_2}\alpha_3(\bar{\Omega}_1)c_{12}+\frac{2\xi}{\bar{\Omega}_2}\alpha_4(\bar{\Omega}_1)c_{14}+\xi^2\alpha_2(\bar{\Omega}_2)\bigg)\\
\dot{c}_8&=&-\bar{\Omega}_2^2c_7\\
\dot{c}_9&=&c_{10}+\frac{1}{1+\xi^2}\bigg(\frac{2\xi}{\bar{\Omega}_2}\alpha_3(\bar{\Omega}_1)c_4+\frac{2\xi}{\bar{\Omega}_2}\alpha_4(\bar{\Omega}_1)c_5\nonumber\\
& &\hspace{1.5cm}-2\xi^2\alpha_4(\bar{\Omega}_2)c_9+\frac{2\xi^2}{\bar{\Omega}_2^2}\alpha_3(\bar{\Omega}_2)c_{10}\bigg)\\
\dot{c}_{10}&=&-\bar{\Omega}_2^2 c_9\\
\dot{c}_{11}&=&\frac{1}{1+\xi^2}\left(\frac{4 \xi}{\bar{\Omega}_2}\alpha_3(\bar{\Omega}_1)c_1+\frac{2 \xi}{\bar{\Omega}_2}\alpha_4(\bar{\Omega}_1)c_2\right)\\
& &-\frac{2\xi^2}{1+\xi^2}\alpha_4(\bar{\Omega}_2)c_{11}+\left(1+\frac{2\xi^2}{\bar{\Omega}_2^2(1+\xi^2)}\right)c_{12}\nonumber\\
& &+c_{13}+\frac{\xi}{\bar{\Omega}_2(1+\xi^2)}\alpha_2(\bar{\Omega}_1)\\
\dot{c}_{12}&=&-\bar{\Omega}_2^2c_{11}+c_{14}
\end{eqnarray*}
\begin{eqnarray}
\dot{c}_{13}&=&\frac{1}{1+\xi^2}\bigg(\frac{2\xi}{\bar{\Omega}_2}\alpha_3(\bar{\Omega}_1)c_2+\frac{4\xi}{\bar{\Omega}_2}\alpha_4(\bar{\Omega}_1)c_3+4\xi\bar{\Omega}_2\alpha_4(\bar{\Omega}_2)c_6\bigg)\nonumber\\
& &- \frac{2\xi}{(1+\xi^2)\bar{\Omega}_2} \alpha_3(\bar{\Omega}_2)c_7-\bigg(\frac{2}{1+\xi^2}\alpha_3(\bar{\Omega}_1)+\bar{\Omega}_1^2\bigg)c_{11}\nonumber\\
& &-\frac{2}{1+\xi^2}\bigg(\alpha_4(\bar{\Omega}_1)+\xi^2\alpha_4(\bar{\Omega}_2)\bigg)c_{13}\nonumber\\
& &+\bigg(1+\frac{2\xi^2}{(1+\xi^2)\bar{\Omega}_2^2}\alpha_3(\bar{\Omega}_2)\bigg)c_{14}\nonumber\\
& &-\frac{\xi}{(1+\xi^2)\bar{\Omega}_2}\bigg(\alpha_1(\bar{\Omega}_1)+\alpha_1(\bar{\Omega}_2)\bigg)\nonumber\\
\dot{c}_{14}&=&\frac{1}{1+\xi^2}\left(2\xi\bar{\Omega}_2\alpha_4(\bar{\Omega}_2)c_7-\frac{4\xi}{\bar{\Omega}_2}\alpha_3(\bar{\Omega}_2)c_8\right)\nonumber\\
& &-\bigg(\bar{\Omega}_1^2+\frac{2}{1+\xi^2}\alpha_3(\bar{\Omega}_1)\bigg)c_{12}-\bar{\Omega}_2^2 c_{13}\nonumber\\
& &-\frac{1}{1+\xi^2}\left(2\alpha_4(\bar{\Omega}_1)c_{14}+\xi\bar{\Omega}_2\alpha_2(\bar{\Omega}_2)\right)\label{DGLs}\,.
\end{eqnarray}
The expectation values of the anticommutators can be given in terms of the
functions $c_i(t)$ according to
\begin{eqnarray}
\langle\{\hat{\bar{Q}}_1,\hat{\bar{Q}}_1\}\rangle&=&2(2c_1+c_4^2)\nonumber\\
\langle\{\hat{\bar{Q}}_1,\hat{\bar{Q}}_2\}\rangle&=&2(c_{11}+c_4c_9)\nonumber\\
\langle\{\hat{\bar{Q}}_2,\hat{\bar{Q}}_2\}\rangle&=&2(2c_6+c_9^2)\nonumber\\
\langle\{\hat{\bar{Q}}_1,\hat{\bar{P}}_1\}\rangle&=&2(c_2+c_4c_5)\nonumber\\
\langle\{\hat{\bar{Q}}_2,\hat{\bar{P}}_2\}\rangle&=&2(c_7+c_9c_{10})\nonumber\\
\langle\{\hat{\bar{Q}}_1,\hat{\bar{P}}_2\}\rangle&=&2(c_{12}+c_{4}c_{10})\nonumber\\
\langle\{\hat{\bar{P}}_1,\hat{\bar{P}}_1\}\rangle&=&2(2c_3+c_5^2)\nonumber\\
\langle\{\hat{\bar{P}}_2,\hat{\bar{P}}_2\}\rangle&=&2(2c_8+c_{10}^2)\nonumber\\
\langle\{\hat{\bar{P}}_1,\hat{\bar{P}}_2\}\rangle&=&2(c_{14}+c_5c_{10})\nonumber\\
\langle\{\hat{\bar{Q}}_2,\hat{\bar{P}}_1\}\rangle&=&2(c_{13}+c_5c_9)\label{comm}\,.
\end{eqnarray}

\subsection{Approximate Solutions of the Differential Equations}

From the initial condition (\ref{eq:initial}) we find that the functions $c_4,c_5,c_9$ and $c_{10}$ are
vanishing since they describe momentum and position displacements that are absent in
symmetric Gaussians.
At $t=0$, the non-vanishing anticommutators have the expectation values
\begin{eqnarray*}
\langle\{\hat{p}_{S/A},\hat{p}_{S/A}\}\rangle&=&\frac{1}{\langle\{\hat{q}_{S/A},\hat{q}_{S/A}\}\rangle}=\omega_0\\ \langle\{\hat{P}_{S/A},\hat{P}_{S/A}\}\rangle&=&\frac{1}{\langle\{\hat{Q}_{S/A},\hat{Q}_{S/A}\}\rangle}=\kappa\Omega_0.
\end{eqnarray*}
Neglecting terms of order $g^2\gamma$ in the differential equation, some of the 
equations (\ref{DGLs}) decouple from each other.
For the coefficients concerning the oscillator with the variable $\bar{Q}_1$,
we find
\begin{eqnarray}\label{Approx1}
c_1&=&\sum_{i=1}^3A_i e^{\lambda_i t}+\frac{\alpha_1(\bar{\Omega}_1)-2\alpha_2(\bar{\Omega}_1)\alpha_4(\bar{\Omega}_1)}{4\alpha_4(\bar{\Omega}_1)(2\alpha_3(\bar{\Omega}_1)+\bar{\Omega}_1^2)}\nonumber\\
c_2&=&\sum_{i=1}^3A_i\lambda_i e^{\lambda_i t}\nonumber\\
c_3&=&\sum_{i=1}^3A_i\left(\frac{\lambda_i}{2}+\bar{\Omega}_1^2+2\alpha_3(\bar{\Omega}_1)+\alpha_4(\bar{\Omega}_1)\lambda_i\right)\lambda_i e^{\lambda_i t}\nonumber\\
& &+\frac{\alpha_1(\bar{\Omega}_1)}{4\alpha_4(\bar{\Omega}_1)}\,,
\end{eqnarray}
with the eigenmodes
\begin{eqnarray*}
\lambda_1&=&-2\alpha_4(\bar{\Omega}_1)\\
\lambda_{2,3}&=&-2\left(\alpha_4(\bar{\Omega}_1)\pm i\sqrt{\bar{\Omega}_1^2+2\alpha_3(\bar{\Omega}_1)-\alpha_4^2(\bar{\Omega}_1)}\right)\,.
\end{eqnarray*}
The $A_i$ are chosen such that
\begin{eqnarray*}
c_1(0)&=&\frac{\omega_0+\xi^2\,\kappa\Omega_0 }{4\kappa\Omega_0\omega_0(1+\xi^2)}\\
c_2(0)&=&0\\
c_3(0)&=&\frac{\kappa\Omega_0+\xi^2\omega_0}{4(1+\xi^2)}\,.
\end{eqnarray*}
For the oscillator  $\bar{Q}_2$ we find
\begin{eqnarray}\label{Approx2}
c_6&=&B_1e^{2i\bar{\Omega}_2 t}+B_2 e^{-2 i \bar{\Omega}_2 t}+B_3\nonumber\\
c_7&=&2 i \bar{\Omega}_2 B_1e^{2i\bar{\Omega}_2 t}-2 i \bar{\Omega}_2 B_2e^{-2i\bar{\Omega}_2 t}\nonumber\\
c_8&=&-\bar{\Omega}_2^2(B_1e^{2i\bar{\Omega}_2 t}+B_2 e^{-2 i \bar{\Omega}_2 t}-B_3)\nonumber\,.
\end{eqnarray}
The $B_i$ are chosen such that
\begin{eqnarray*}
c_6(0)&=&\frac{\omega_0+\xi^2 \kappa\Omega_0}{4 \bar{\Omega}_2^2(1+\xi^2)}\\
c_7(0)&=&0\\
c_8(0)&=&\frac{\bar{\Omega}_2^2(\kappa\Omega_0 +\xi^2\omega_0 )}{4\kappa\Omega_0\omega_0(1+\xi^2)}
\end{eqnarray*}
The coupling between $\bar{Q}_1$ and $\bar{Q}_2$ is described by $c_{11}...c_{14}$, which read
\begin{eqnarray}\label{Approx3}
c_{11}&=&\sum_{i=1}^4 C_i e^{\kappa_it}\\
c_{12}&=&-\sum_{i=1}^4C_i \frac{2 \bar{\Omega}_2^2 (\alpha_4(\bar{\Omega}_1)+\kappa_i)}{2\alpha_3(\bar{\Omega}_1)+\bar{\Omega}_1^2-\bar{\Omega}_2^2+2\alpha_4(\bar{\Omega}_1)\kappa_i+\kappa_i^2}e^{\kappa_it}\nonumber\\
c_{13}&=&\sum_{i=1}^4C_i\kappa_ie^{\kappa_it}\nonumber\\
& &+\sum_{i=1}^4C_i\frac{2 \bar{\Omega}_2^2(\alpha_4(\bar{\Omega}_1)+\kappa_i)}{2\alpha_3(\bar{\Omega}_1)+\bar{\Omega}_1^2-\bar{\Omega}_2^2+2\alpha_4(\bar{\Omega}_1)\kappa_i+\kappa_i^2}e^{\kappa_it}\nonumber\\
c_{14}&=&\sum_{i=1}^4C_i\frac{\bar{\Omega}_2^2 (2\alpha_3(\bar{\Omega}_1)+\bar{\Omega}_1^2-\bar{\Omega}_2^2-\kappa_i^2)}{2\alpha_3(\bar{\Omega}_1)+\bar{\Omega}_1^2-\bar{\Omega}_2^2+2\alpha_4(\bar{\Omega}_1)\kappa_i+\kappa_i^2}e^{\kappa_it}\nonumber\,,
\end{eqnarray}
with the eigenmodes
\begin{eqnarray*}
\kappa_{1,2,3,4}&=&-\alpha_4(\bar{\Omega}_1)\pm i\bigg(2 \alpha_3(\bar{\Omega}_1)- 
    \alpha_4^2(\bar{\Omega}_1) + \bar{\Omega}_1^2 + \bar{\Omega}_2^2\nonumber\\
&\pm& 2 \bar{\Omega}_2\sqrt{2 \alpha_3(\bar{\Omega}_1) - \alpha_4^2(\bar{\Omega}_1) + \bar{\Omega}_1^2}\bigg)^{1/2}\,.
\end{eqnarray*}
The $C_i$ are chosen such that
\begin{eqnarray*}
c_{11}(0)&=&0\\
c_{12}(0)&=&\frac{\xi\,\bar{\Omega}_2(\omega_0-\kappa\Omega_0)}{2\kappa\Omega_0\omega_0(1+\xi^2)}\\
c_{13}(0)&=&\frac{\xi(\omega_0-\kappa\Omega_0)}{2\bar{\Omega}_2(1+\xi^2)}\\
c_{14}(0)&=&0\,.
\end{eqnarray*}

\subsection{Bath Correlators}\label{bathcorr}
For the decoherence rates we find
\begin{eqnarray}\label{corr1}
\alpha_{1,S/A}(\bar{\Omega})=\frac{\pi}{4} J^{3D}(\bar{\Omega})\left(1\pm\frac{\sin(\bar{\Omega}r)}{\bar{\Omega}r}\right)
\coth\left(\frac{\bar{\Omega}}{2 T}\right)\,.
\end{eqnarray}
The dissipation correlators read
\begin{eqnarray}\label{corr2}
\alpha_{4,S/A}(\bar{\Omega})=\frac{\pi J^{3D}(\bar{\Omega})}{4\bar{\Omega}}\left(1\pm\frac{\sin(\bar{\Omega}r)}{\bar{\Omega}r}\right)\,.
\end{eqnarray}
The anomalous diffusion correlators have the form
\begin{eqnarray}\label{corr3}
\alpha_{2,S/A}(\bar{\Omega})&=&\frac{2\gamma \Gamma(1+s)}{\pi}\left(\frac{\bar{\Omega}}{\Omega_c}\right)^{s-1}\coth\left(\frac{\bar{\Omega}}{2 T}\right)\times\nonumber\\
& &\hspace{-2cm}\times\Bigg[e^{-\frac{\bar{\Omega}}{\Omega_c}}\bigg(\cos(\pi s)\Re\left\{\Gamma\left(-s,-\frac{\bar{\Omega}}{\Omega_c}\right)\right\}\nonumber\\
& &\hspace{-2cm}-\sin(\pi s)\Im\left\{\Gamma\left(-s,-\frac{\bar{\Omega}}{\Omega_c}\right)\right\}\bigg)+e^{\frac{\bar{\Omega}}{\Omega_c}}\Gamma\left(-s,\frac{\bar{\Omega}}{\Omega_c}\right)\Bigg]\nonumber\\
& &\hspace{-2cm}-\frac{8 T\gamma\Omega_c \Gamma(1+s)}{\pi}\left(\frac{1}{\bar{\Omega}^2 s}+\sum_{n=1}^\infty \left(\frac{2\pi n T}{\Omega_c}\right)^{s}\frac{c_n}{\bar{\Omega}^2+(2\pi n T)^2}\right)\nonumber\\
& &\hspace{-2cm}\pm \Bigg\{-\frac{ \gamma \Gamma(s-1)}{\pi r \Omega_c}\left(\frac{\bar{\Omega}}{\Omega_c}\right)^{s-2}\coth\left(\frac{\bar{\Omega}}{2 T}\right)\times\nonumber\\
& &\hspace{-2cm}\times\Bigg[2\cos(\bar{\Omega}r)e^{-\frac{\bar{\Omega}}{\Omega_c}}\bigg(\sin(\pi s)\Re\left\{\Gamma\left(-s+2,-\frac{\bar{\Omega}}{\Omega_c}\right)\right\}\nonumber\\
& &\hspace{-2cm}+\cos(\pi s)\Im\left\{\Gamma\left(-s+2,-\frac{\bar{\Omega}}{\Omega_c}\right)\right\}\bigg)\nonumber\\
& &\hspace{-2cm}+i e^{\bar{\Omega}\left(\frac{1}{\Omega_c}+i r\right)}\Gamma\left(-s+2,\bar{\Omega}\left(\frac{1}{\Omega_c}+i r\right)\right)\nonumber\\
& &\hspace{-2cm}-i e^{\bar{\Omega}\left(\frac{1}{\Omega_c}-i r\right)}\Gamma\left(-s+2,\bar{\Omega}\left(\frac{1}{\Omega_c}-i r\right)\right)\nonumber\\
& &\hspace{-2cm}-i e^{-\bar{\Omega}\left(\frac{1}{\Omega_c}+i r\right)-i\pi s}\Gamma\left(-s+2,-\bar{\Omega}\left(\frac{1}{\Omega_c}+i r\right)\right)\nonumber\\
& &\hspace{-2cm}+i e^{-\bar{\Omega}\left(\frac{1}{\Omega_c}-i r\right)+i \pi s}\Gamma\left(-s+2,-\bar{\Omega}\left(\frac{1}{\Omega_c}-i r\right)\right) \Bigg]\nonumber\\
& &\hspace{-2cm}+\frac{4\gamma T\Gamma(s-1)}{\pi r}\sum_{n=1}^\infty \left(\frac{2\pi n T}{\Omega_c}\right)^{s-1}\frac{d_n}{\bar{\Omega}^2+(2\pi n T)^2}\Bigg\}
\end{eqnarray}
with
\begin{eqnarray*}
c_n&=&e^{-i\frac{2\pi n T}{\Omega_c}-i\frac{\pi s}{2}}\Gamma\left(-s,-i\frac{2\pi n T}{\Omega_c}\right)\\
& +&e^{i\frac{2\pi n T}{\Omega_c}+i\frac{\pi s}{2}}\Gamma\left(-s,i\frac{2\pi n T}{\Omega_c}\right)\,;\nonumber\\
d_n&=&e^{i 2\pi n T\left(\frac{1}{\Omega_c}+i r\right)+i\frac{\pi s }{2}}\Gamma\left(-s+2,i 2\pi n T\left(\frac{1}{\Omega_c}+i r\right)\right)\nonumber\\
& +&e^{-i 2\pi n T\left(\frac{1}{\Omega_c}-i r\right)-i\frac{\pi s }{2}}\Gamma\left(-s+2,-i 2\pi n T\left(\frac{1}{\Omega_c}-i r\right)\right)\nonumber\\
& -&e^{-i 2\pi n T\left(\frac{1}{\Omega_c}+i r\right)-i\frac{\pi s }{2}}\Gamma\left(-s+2,-i 2\pi n T\left(\frac{1}{\Omega_c}+i r\right)\right)\nonumber\\
& +&e^{i 2\pi n T\left(\frac{1}{\Omega_c}-i r\right)+i\frac{\pi s }{2}}\Gamma\left(-s+2,i 2\pi n T\left(\frac{1}{\Omega_c}-i r\right)\right)\,.
\end{eqnarray*}
The Lamb shifts due to the bath correlators are given by
\begin{eqnarray}\label{corr4}
\alpha_{3,S/A}(\bar{\Omega})&=&-\frac{2\gamma \Gamma(3+s)\bar{\Omega}}{\pi}\left(\frac{\bar{\Omega}}{\Omega_c}\right)^{s-1}\times\nonumber\\
& &\hspace{-2cm}\times
\Bigg[e^{-\frac{\bar{\Omega}}{\Omega_c}}\bigg(\cos(\pi s)\Re\left\{\Gamma\left(-2-s,-\frac{\bar{\Omega}}{\Omega_c}\right)\right\}\nonumber\\
& &\hspace{-2cm}-\sin(\pi s)\Im\left\{\Gamma\left(-2-s,-\frac{\bar{\Omega}}{\Omega_c}\right)\right\}\bigg)\nonumber\\
& &\hspace{-2cm}+e^{\frac{\bar{\Omega}}{\Omega_c}}\Gamma\left(-2-s,\frac{\bar{\Omega}}{\Omega_c}\right)\Bigg]+\frac{4\gamma\Omega_c^3 \Gamma(3+s)}{\pi\bar{\Omega}^2 (2+s)}\nonumber\\
& &\hspace{-2cm}\pm\Bigg\{\frac{ \gamma \Gamma(1+s)}{\pi r }\left(\frac{\bar{\Omega}}{\Omega_c}\right)^{s-1}\times\nonumber\\
& &\hspace{-2cm}\times\Bigg[2\cos(\bar{\Omega}r)e^{-\frac{\bar{\Omega}}{\Omega_c}}\bigg(\sin(\pi s)\Re\left\{\Gamma\left(-s,-\frac{\bar{\Omega}}{\Omega_c}\right)\right\}\nonumber\\
& &\hspace{-2cm}+\cos(\pi s)\Im\left\{\Gamma\left(-s,-\frac{\bar{\Omega}}{\Omega_c}\right)\right\}\bigg)\nonumber\\
& &\hspace{-2cm}+i e^{\bar{\Omega}\left(\frac{1}{\Omega_c}+i r\right)}\Gamma\left(-s,\bar{\Omega}\left(\frac{1}{\Omega_c}+i r\right)\right)\nonumber\\
& &\hspace{-2cm}-i e^{\bar{\Omega}\left(\frac{1}{\Omega_c}-i r\right)}\Gamma\left(-s,\bar{\Omega}\left(\frac{1}{\Omega_c}-i r\right)\right)\nonumber\\
& &\hspace{-2cm}-i e^{-\bar{\Omega}\left(\frac{1}{\Omega_c}+i r\right)-i\pi s}\Gamma\left(-s,-\bar{\Omega}\left(\frac{1}{\Omega_c}+i r\right)\right)\nonumber\\
& &\hspace{-2cm}+i e^{-\bar{\Omega}\left(\frac{1}{\Omega_c}-i r\right)+i \pi s}\Gamma\left(-s,-\bar{\Omega}\left(\frac{1}{\Omega_c}-i r\right)\right) \Bigg]\Bigg\}\,.
\end{eqnarray}
The Lamb shifts originating from the counter terms are determined by
\begin{eqnarray}
& &\int_0^\infty d\omega \frac{J^{3D}(\omega)}{\omega}\left(1\pm\frac{\sin(\omega r )}{\omega r}\right)=\frac{8\gamma \Omega_c}{\pi}\bigg(\Gamma(s)\\
& &\pm\frac{(1+\Omega_c^2 r^2)^{-\frac{s-1}{2}}\Gamma(s-1)\sin\left[(s-1)\arctan(\Omega_c r)\right]}{\Omega_c r}\bigg)\,.\nonumber
\end{eqnarray}


\begin{thebibliography}{99}
 

\bibitem{JZKGKS03}
E.~Joos et al., \textit{Decoherence and the Appearance of a Classical World in Quantum Theory}, Springer (2003).

\bibitem{S07}
M.~Schlosshauer, \textit{Decoherence and the Quantum-to-Classical Transition},
Springer (2007).

\bibitem{Bra02}
D.~Braun, Phys.~Rev.~Lett.~{\bf 89}, 277901 (2002).

\bibitem{BFP03}
F.~Benatti, R.~Floreanini and M.~Piani, Phys.~Rev.~Lett.~{\bf 91}, 070402 (2003).

\bibitem{Bra05}
D.~Braun, Daniel, Phys.~Rev.~A {\bf 72}, 062324 (2005).

\bibitem{BF06}
F.~Benatti and R.~Floreanini, J.~Phys.~A {\bf 39}, 2689 (2006).

\bibitem{OK06}
S.~Oh and J.~Kim, Phys.~Rev.~A {\bf 73}, 062306 (2006).

\bibitem{Pra04}
J.~S.~Prauzner-Bechcicki, J.~Phys.~A, {\bf 37}, L173 (2004).

\bibitem{AZ07}
J.~H.~An and W.~M.~Zhang, Phys.~Rev.~A {\bf 76}, 042127 (2007).

\bibitem{CYH08}
C.-H. Chou, T.~Yu and B.~L.~Hu, Phys.~Rev.~E {\bf 77}, 011112 (2008).

\bibitem{LG07}
K.-L.~Liu and H.-S.~Goan, Phys.~Rev.~A {\bf 76}, 022312 (2007).

\bibitem{HB08}
C.~H\"orhammer and H.~B\"uttner, Phys.~Rev.~A {\bf 77}, 042305 (2008).

\bibitem{PR08}
J.~P.~Paz and A.~J.~Roncaglia, Phys.~Rev.~Lett.~{\bf 100}, 220401 (2008).

\bibitem{STP07}
D.~Solenov, D.~Tolkunov and V.~Privman, Phys.~Rev.~B {\bf 75}, 035134 (2007).

\bibitem{ZQK09}
T.~Zell, F.~Queisser and R.~Klesse, Phys.~Rev.~Lett.~{\bf 102}, 160501 (2009).

\bibitem{UZ89}
W.~G.~Unruh and W.~H.~Zurek, Phys.~Rev.~D, {\bf 40}, 1071 (1989).

\bibitem{WW08}
J.~Wilkie and Y.~M.~Wong, J.~Phys.~A {\bf 41}, 335005 (2008).
 
\bibitem{VW02}
G.~Vidal and R.~F.~Werner, Phys.~Rev.~A {\bf 65}, 032314 (2002).

\bibitem{GZ04}
C.~W.~Gardiner and P.~Zoller, \textit{Quantum Noise}, Springer (2004).


\end{thebibliography}
\end{document}